\documentclass[%
 reprint,
 amsmath,amssymb,
 aps,
]{revtex4-2}

\usepackage{graphicx}
\usepackage{dcolumn}
\usepackage{bm}

\begin{document}

\title{Whispering gallery quantum well exciton polaritons in an Indium Gallium Arsenide microdisk cavity}%

\author{Romain de Oliveira }
\thanks{These authors contributed equally}
 \affiliation{Matériaux et Phénomènes Quantiques, Université Paris Cité, CNRS UMR 7162, 10 rue Alice Domon et Léonie Duquet 75013 Paris, France}
 
\author{Martin Colombano}
\thanks{These authors contributed equally}
\affiliation{Matériaux et Phénomènes Quantiques, Université Paris Cité, CNRS UMR 7162, 10 rue Alice Domon et Léonie Duquet 75013 Paris, France}

\author{Florent Malabat}
\affiliation{Matériaux et Phénomènes Quantiques, Université Paris Cité, CNRS UMR 7162, 10 rue Alice Domon et Léonie Duquet 75013 Paris, France}

\author{Martina Morassi}
\affiliation{Centre de Nanosciences et Nanotechnologies, CNRS UMR 9001, Université Paris-Saclay, 91120 Palaiseau, France}
\author{Aristide Lemaître}
\affiliation{Centre de Nanosciences et Nanotechnologies, CNRS UMR 9001, Université Paris-Saclay, 91120 Palaiseau, France}
\author{Ivan Favero}
\email{ivan.favero@u-paris.fr} 
\affiliation{Matériaux et Phénomènes Quantiques, Université Paris Cité, CNRS UMR 7162, 10 rue Alice Domon et Léonie Duquet 75013 Paris, France}

\begin{abstract}
Despite appealing high-symmetry properties that enable high quality factor and strong confinement, whispering gallery modes of spherical and circular resonators have been absent from the field of quantum-well exciton polaritons. Here we observe whispering gallery exciton polaritons in a Gallium Arsenide microdisk cavity filled with Indium Gallium Arsenide quantum wells, the testbed materials of polaritonics. Strong coupling is evidenced in photoluminescence and resonant spectroscopy, accessed through concomitant confocal microscopy and near-field optical techniques. Excitonic and optical resonances are tuned by varying temperature and disk radius, revealing Rabi splittings between 5 and 10 meV. A dedicated analytical quantum model for such circular polaritons is developed, which reproduces the measured values. At high power, lasing is observed and accompanied by a blueshift of the emission that points to the regime of polariton lasing.
\end{abstract}

\maketitle


\section{\label{sec:level1}Introduction}

Exciton-polaritons are half-light half-matter particles, resulting from the strong coupling between photons and excitons. Ever since the pioneering work of Hopfield on bulk semiconductor polaritons \cite{hopfield} and their first experimental observation in a semiconductor quantum well \cite{weisbuch}, the field of polaritonics has experienced spectacular developments in nonlinear optics \cite{ciuti,ferrier} and quantum many-body physics, culminating with the observation of non-equilibrium Bose-Einstein condensates \cite{kasprzak,balili,bajoni2008} and superfluid phenomena \cite{amo,lagoudakis}. Most of these achievements were enabled by coupling quantum well excitons to planar (2D) \cite{kasprzak,balili}, linear (1D) \cite{wertz}, and pillar (0D) \cite{bajoni2008} optical cavities (Fig. \ref{fig:1} (a)). Along these scientific advances, Indium Arsenide (InAs) and Gallium Arsenide (GaAs), together with their ternary alloys, have often served as testbed materials.

In parallel, an original approach was followed in large bandgap semiconductors vertically grown in the form of hexagonal microwires, where the bulk exciton naturally coupled to the hexagonal optical gallery modes of the wire. Implemented both in Zinc Oxyde \cite{Shen2008,Dang2011} and Gallium Nitride \cite{Richard2012,Cho2015}, it enabled reaching large Rabi splitting. Yet with their low hexagonal symmetry, these modes rather ressemble Mie resonances: they strongly emit in the far-field and present poor temporal confinement, with a quality factor that saturates in the hundred range. This strongly contrasts with whispering gallery modes (WGMs) of circular and spherical objects, which enable long temporal confinement, with basically no emission in the far field. The ultra-high quality factors of circular and spherical WGMs impacted many fields such as microlasers \cite{mccall}, sensing \cite{vollmer,zhu}, optomechanics \cite{schliesser,ding}, non-linear optics \cite{strekalov,roland} and cavity quantum electrodynamics \cite{vernooy}. The quality factor of low refractive index material WGMs attains $10^{10}$ in millimeter-scale circular resonators \cite{malekiQ}, while WGMs in high refractive index GaAs have reached six million for a wavelength-sized disk cavity \cite{guha}. In GaAs material, it was theoretically envisioned that coupling the bulk exciton to such WGMs of a submicron sphere would lead to superior light-matter interactions \cite{Kavokin2009}. The experiment was however never realized, be it with the bulk exciton or with a quantum-well exciton. Amongst different reasons, one difficulty relies precisely in the highly confined nature of high-Q WGMs, which demands near-field optical techniques for optimal spectroscopy and control. In consequence, the field of quantum-well polaritons did not benefit yet from the superior features of spherical and circular WGMs.

Here we observe the strong coupling between the excitonic transition of an InGaAs/GaAs multiple quantum well (MQW) and the circular whispering gallery modes of a GaAs disk microcavity. Strong coupling is evidenced through a combination of confocal microscopy and near-field techniques, enabling acquiring information from both photoluminescence (PL) and resonant optical spectroscopy of the resonators. Excitonic and optical resonances are tuned in a dual manner by varying the temperature and the disk radius, revealing anti-crossings on multiple polariton branches with Rabi splitting evolving between 5 and 10 meV. We develop a dedicated analytical quantum model to depict these new kind of circular polaritons, which we employ to compute Rabi energies and obtain agreement with experiments with minimal parameters. At high optical power a lasing phenomenon is observed, with a pronounced blueshift and line narrowing of the emission in concordance with the threshold. The observed energy scales are compliant with the regime of polariton lasing, or non-equilibrium condensation of polaritons. The present work hence introduces and validates circular polaritons in WGM disk resonators as a new viable physical platform for light-matter research. Amongst several possible applications, one is already envisioned in hybrid optomechanics, where the mechanical motion of a disk resonator is expected to strongly couple to such polaritons \cite{carlon2022enhanced}.

\section{\label{sec:level2}Device and methods}

\begin{figure}[h!]
\centering
     \includegraphics[width=0.5\textwidth]{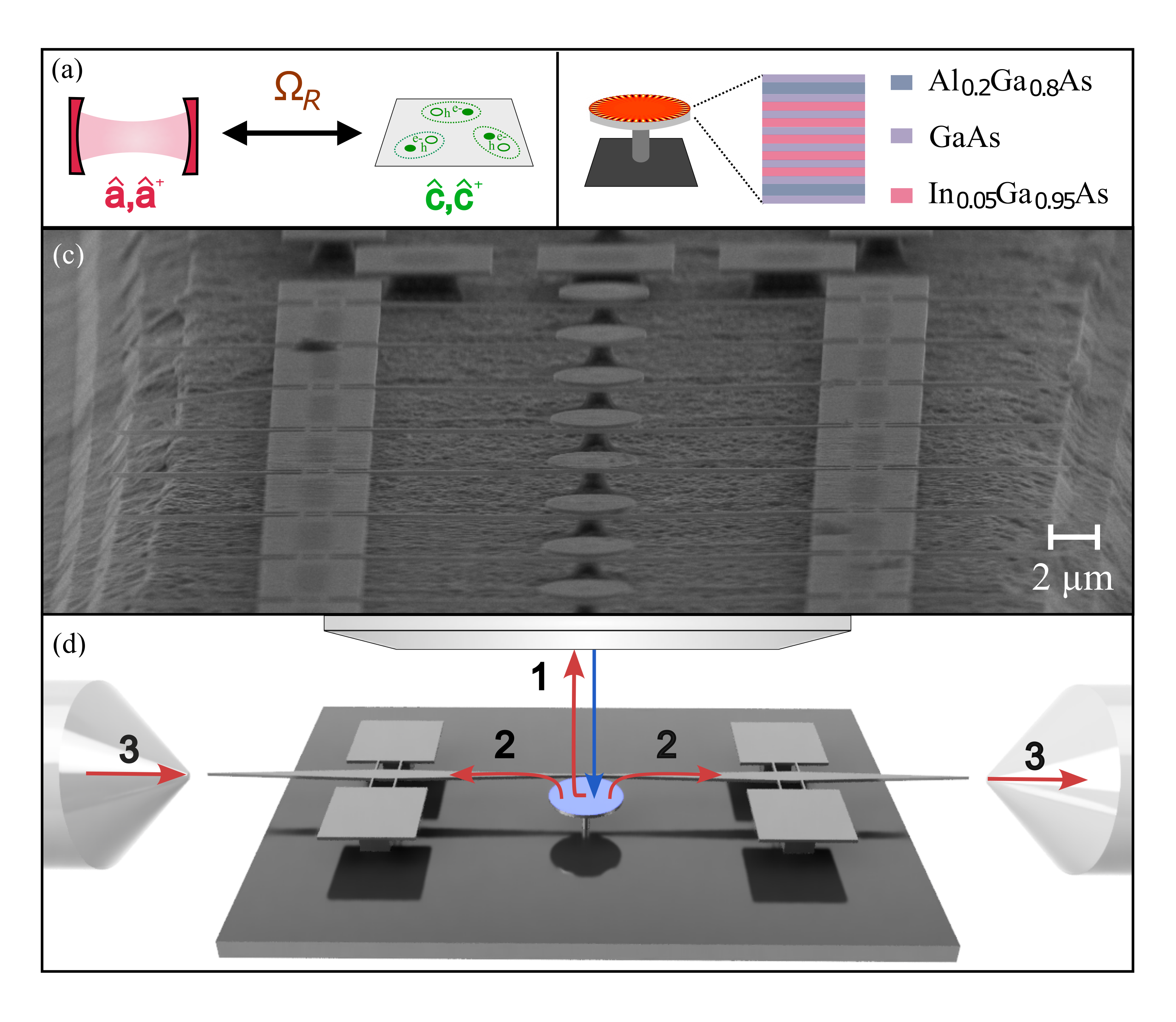}
    \caption{(a) Coupling between the mode of an optical cavity (red) and the excitons of a quasi-planar quantum well (green). (b) Layout of a disk resonator supporting a WGM, and of the MQW structure within the disk layer.  (c) SEM micrograph of the sample containing eight disks and coupling waveguides. Each waveguide suspended by its square anchoring pads enables evanescent coupling to an individual disk, and is terminated by inverted tapers. The latter are protruding over the cliffs of the mesa (extreme left and right of the image). (d) 3D rendering of the sample illustrating the three experimental configurations of this work:  1- Confocal micro-photoluminescence ($\mu$PL) under non-resonant excitation, 2- Non-resonant excitation through confocal objective and emission collection through waveguide, 3- In-plane resonant laser spectroscopy through waveguide.}
    \label{fig:1}
\end{figure}

Five identical 12 nm thick $\text{In}_{0.05}\text{Ga}_{0.95}\text{As}$ QWs, with 15 nm well-to-well separation, are inserted by epitaxy within a GaAs matrix and bounded by two 20 nm thick $\text{Al}_{0.2}\text{Ga}_{0.8}\text{As}$ diffusion barriers, for a total thickness of 200 nm for the top epitaxial layer (see Fig. \ref{fig:1} (b)). The latter is grown over a sacrificial 1800 nm thick $\text{Al}_{0.8}\text{Ga}_{0.2}\text{As}$, sitting on the underlying GaAs substrate. Disks and adjacent tapered waveguides are patterned using e-beam lithography and inductively coupled plasma etching. The sample contains a distribution of disks of varying radius, centered around 2 $\mu$m and varying by steps of 1 nm. The disk pedestals are created using a selective hydrofluoric acid wet etching of the sacrificial AlGaAs, which lets the waveguides freely suspended on their square-shaped anchors. A mesa structure is then created by optical lithography and phosphoric acid wet etching, in order to elevate both disks and waveguides over the rest of the sample. An electron micrograph of the obtained devices is shown in Fig. \ref{fig:1} (c). The mesa enables approaching micro-lensed fibers to inject (collect) light into (from) the inverted tapers, opening the possibility to use the waveguide as a spatially-filtering collecting antenna for the disk emission, or as bus coupler for resonant laser spectroscopy of the disk. Measurements are performed at cryogenic temperatures (4 to 60 K) using a custom cryostat including a cold microscope objective positioned over the sample, providing stable confocal access to the sample with a $2$ $\mu$m radius spot. With a tunable continuous-wave Ti:Sapphire input laser, the set-up offers three complementary experimental configurations to probe the resonators, under resonant or non-resonant excitation, in the far-field or in the near-field (see Fig. \ref{fig:1} (d)). \\

\begin{figure}[h!]
\centering
     \includegraphics[width=0.4\textwidth]{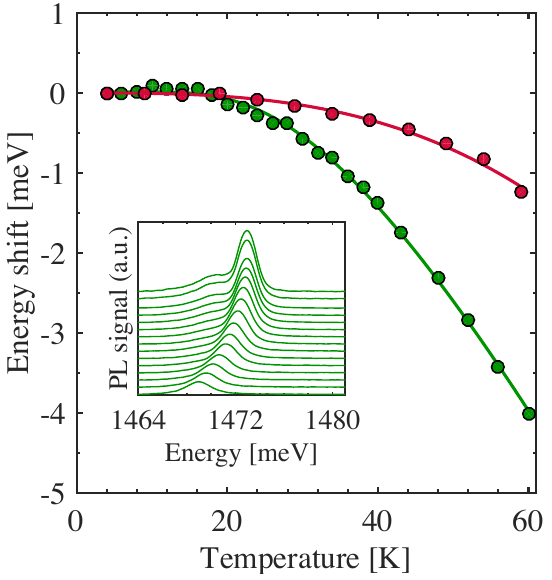}
    \caption{Energy shift of the exciton ($E_x$ - green) and WGMs ($E_c$ - red) as function of temperature. Dots: experiments - Green Solid line: Viña's Model \cite{vina} ($r^2\simeq 0.999$) - Red Solid line: custom minimal model, see text ($r^2 \simeq 0.992$). Inset: Confocal $\mu$PL spectra taken at different temperature from 4 to 60K, and shifted for clarity.}
\label{fig:2}
\end{figure}

In order to investigate the exciton-photon coupling, we must control the energy detuning between the WGM cavity and exciton modes $\delta=E_{c}-E_{x}$. In the following experiments, the detuning is varied by changing the sample temperature or the disk radius. For unambiguous interpretation, we need first to know the temperature dependence of $E_{x}$ and $E_{c}$ in absence of exciton-photon coupling. The first dependance is measured in experimental configuration 1 using confocal photoluminescence, at a location of the sample where there is no disk, hence no WGMs. The second dependance is measured by resonant disk spectroscopy in experimental configuration 3, at a wavelength far enough from the exciton resonance to prevent exciton-WGM coupling. The results are shown in Fig. \ref{fig:2}. The evolution of the MQW exciton energy with temperature (green line) is in agreement with the Viña's Model \cite{vina} $E_{x}(T)=E_{B}-a_{B}(1+2/(\exp^{\Theta_{b}/T}-1))$, while the redshift of the WGM (red line) is fitted with a minimal model $E_{c}(T)=E_{c}(0)\times n_{e}/(n_{e}+A_{1}T+A_{2}T^{2}+A_{3}T^{3})$, with $n_{e}$ the zero-temperature effective refractive index of the disk slab. In the effective index approximation, this dispersion relation is derived by neglecting the evanescent field of high-Q WGMs and the thermal expansion of the disk, while expressing a temperature dependance of the index. This approach leads to a very satisfactory fit of data with only three parameters $A_{i}$, $n_{e}$ being calculated independently. Regarding tuning with the disk radius, increasing the radius $R$ by 1 nm shifts the WGM energy by about 0.3 meV. In order to average out the effects of finite nanofabrication resolution, the sample contains a distribution of disks with nominal radius varying by step of 1 nm and covering a global interval of 40 nm. This allows us working with the same set of WGMs but with a varying initial detuning $\delta_{0}=\delta(T=4K)$. 

\section{\label{sec:level3}Results and Hopfield modeling}
\begin{figure}[h!]
\centering
     \includegraphics[width=0.5 \textwidth]{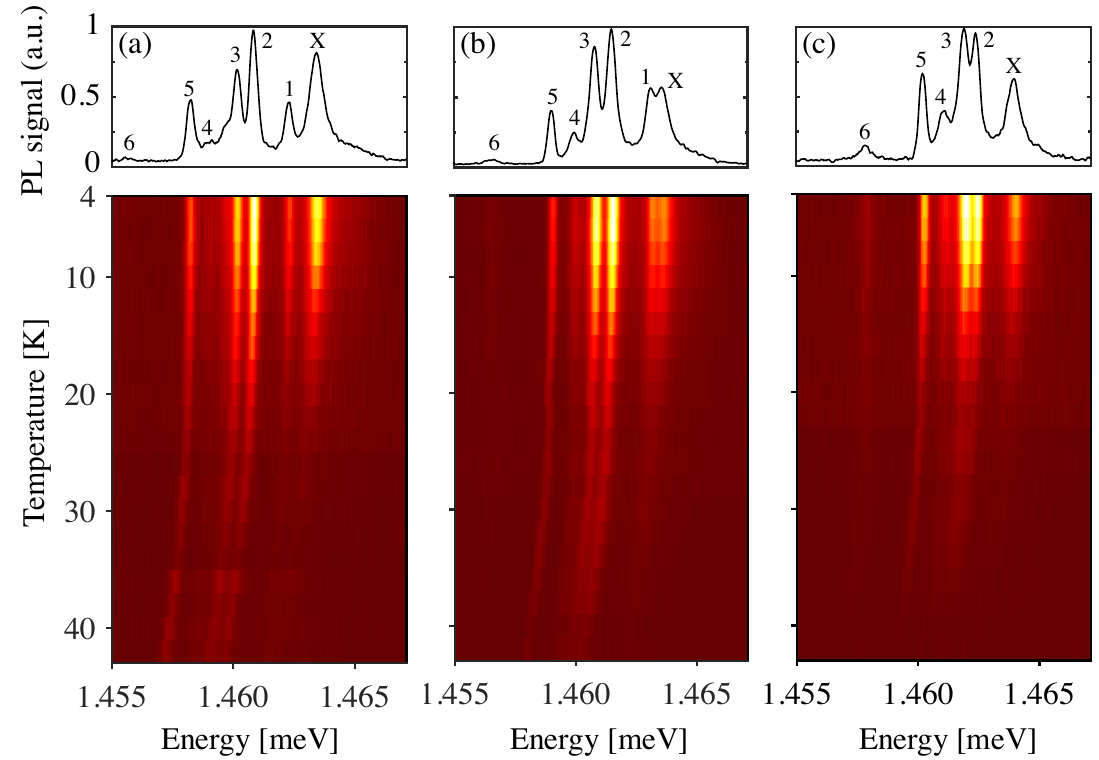}
    
    \caption{PL spectra as a function of temperature for three different disks of decreasing radius: (a) $R=1.992\ \mu m$ (b) $R=1.989\ \mu m$, (c) $R=1.984\ \mu m$. Each spectrum displays the exciton signal (labelled as $X$) as well as several lower-polariton branches labelled with numbers. The temperature starts at 4K and is then incremented in steps of 2K.}
\label{fig:3}
\end{figure}

Fig. \ref{fig:3} represents PL temperature maps for three disks of decreasing radius measured in experimental configuration 2. The first spectrum (top of the map) is taken at 4K, and the temperature is then elevated in steps of 2K. Each individual PL spectrum (horizontal cut of a map) exhibits several peaks: the exciton line (labeled as X), and six lower polariton (LP) lines labeled in order of increasing detuning to the exciton. The concomitant presence of the exciton emission with that LPs is standardly observed on micropillar resonators \cite{bajoni2008}: in etched microstructures some excitons close to surfaces couple less efficiently to confined optical modes and emit directly in free space, despite the system being in the strong coupling regime. In our collected spectrum, each LP branch corresponds to a distinct WGM. When the disk radius decreases, the WGM energies and $\delta_{0}$ increase, producing a blueshift of the different LPs lines as observed when moving from map (a) to (c). In map (c), the LP1 branch seems to disappear: indeed for an increasing detuning LPs adopt a more pronounced exciton character and their PL signal merges with that of the exciton. In our experiments, the PL signal of upper polaritons (UPs) is not observed, which is also a common situation for deeply etched structures \cite{bajoni2008,ferrier,bloch1998}. Upper polaritons tend indeed to relax towards lower energy states available, hence their PL signal is always fainter or undetected. Additionally in our experiment, the UP's PL is collected by the waveguides where it can get absorbed by the QW continuum. Hence below we rather employ the anticrossing of LP lines with the exciton line as a mean to investigate the physics of strong exciton-photon coupling.
\\

\begin{figure}
\centering
\includegraphics[width=0.5\textwidth]{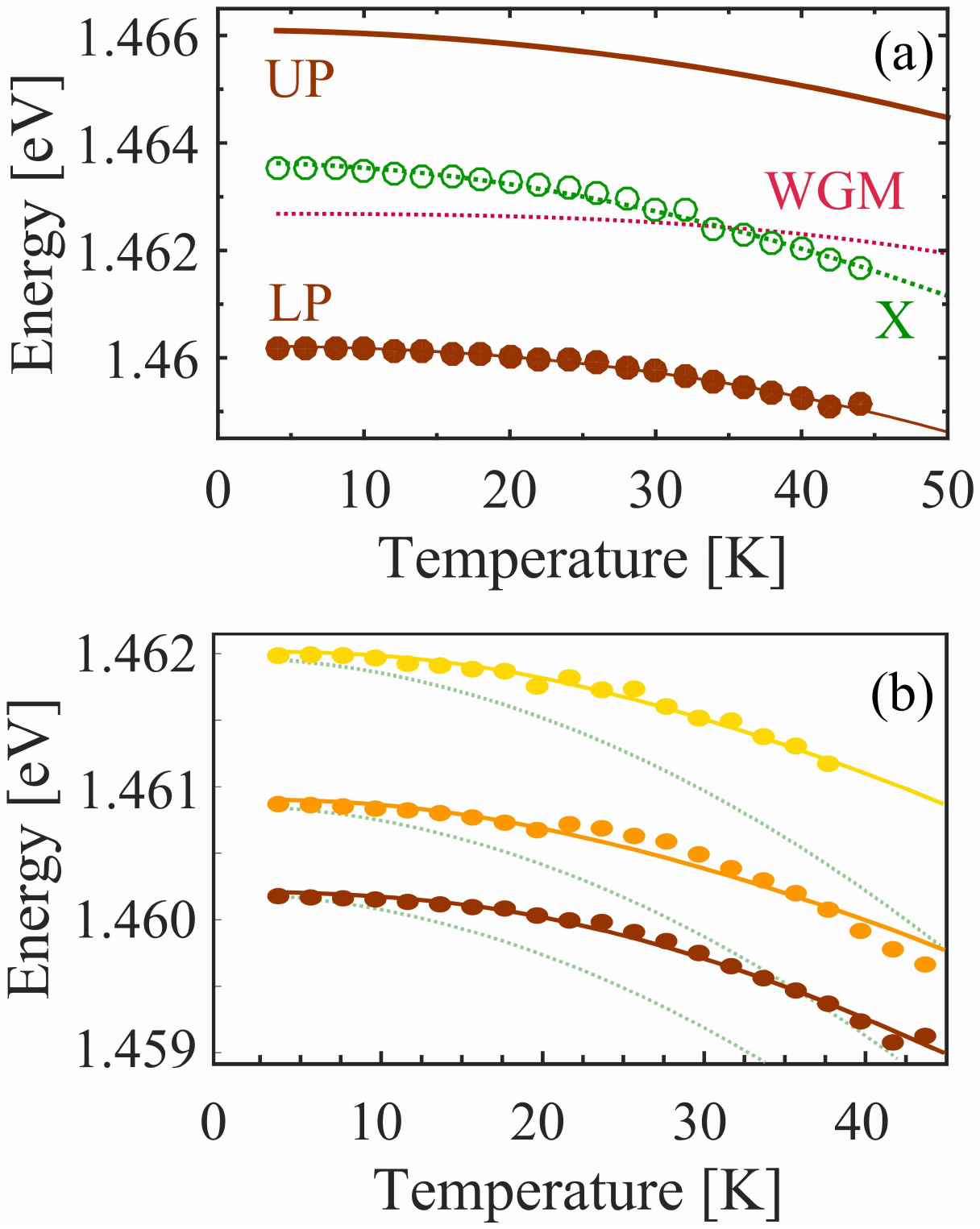}
    \caption{(a) Energy dispersion as a function of temperature for one branch (LP3) of a given disk ($R=1.992 \ \mu m$). Marron circles: measured LP3 emission energy; Open green circles: measured exciton emission energy; Solid marron lines: polariton energy dispersion derived from the Hopfield model; Dashed red line: WGM energy obtained through $\delta_{0}$ obtained from the fitting procedure and the priorly measured temperature evolution of the WGM. (b) LP3 PL energy as function of temperature for three disks of decreasing radius (Marron: $R=1.992 \ \mu m$ - Orange: $R=1.989 \ \mu m$ - Yellow: $R= 1.984 \ \mu m$). Circles: experiments; Solid lines: LP branches predicted by the Hopfield model. Green dashed lines: bare exciton dispersion curves, represented with an offset on each LP branch.}
\label{fig:4}
\end{figure}

We use the Hopfield model to reproduce the observed polaritonic energy spectra with their different branches. The exciton energy is always measurable, hence always known. The Hopfield model relies hence for each WGM on two parameters: the Rabi coupling $\Omega_{R}$ and the detuning $\delta$. Since the temperature dependance of $\delta$ has been measured before, and since the Rabi coupling does not depend on temperature, the mere knowledge of $\delta_{0}$ and $\Omega_{R}$ suffices to predict the temperature evolution of the energy for a given polariton branch. However for each branch, we find that several couples of values of $(\Omega_{R},\delta_{0})$ can lead to a satisfactory fit of the measured temperature evolution. To lift this ambiguity, we carry the fitting procedure in parallel on several disks of varying radius, hence varying $\delta_{0}$, assuming that $\Omega_{R}$ remains constant for the considered radius variations of a few nanometers. With this approach, the fitting procedure returns a unique value of $\Omega_{R}$ and $\delta_{0}$ for each branch, on each disk. Figure \ref{fig:4} represents the results of this parallel fitting procedure for the LP3 branch, tracked as function of temperature on a set of five different disks of distinct radius (only three are displayed for readability). Figure \ref{fig:4} (a) focuses on one of the disks and shows the measured exciton (green open circles) and LP (marron circles), together with the WGM energy and the predictions of the Hopfield model (solid marron lines) for the lower and upper polaritons. A clear anti-crossing behavior is visible over the investigated temperature range, while the model perfectly reproduces the dispersion of the LP with a value of $\delta_{0}=-0.9435$ meV and $\Omega_{R}=5.79$ meV. Figure \ref{fig:4} (b) displays the result of the parallel fitting procedure on LP3 for three different disks of the set, showing the quality of the agreement between experimental data and model ($r^2 > 0.95$). On this panel, an offset exciton line (dashed green) is also represented close to each LP branch, in order to better appreciate the change in dispersion induced by the strong coupling regime. Data for other LP branches are shown in the supplements \cite{supp}.

\section{\label{sec:level4}Theoretical model for Rabi splitting}

In parallel of the above experimental analysis providing a measurement of the Rabi energy, we develop a new model to independently calculate $\Omega_{R}$ with minimal assumptions (see supplements \cite{supp}). For circular QW polaritons, such model was indeed absent from the literature. The QW selection rules impose that the interband heavy hole transition only couple to the transverse electric (TE) WGMs (magnetic field perpendicular to the disk plane $H_{z}$). We restrict the analysis to this polarization and use the effective index method to separate variables for the WGM field $H_{z}(\boldsymbol{r})=\Psi(r)\Lambda(\theta)\chi(z)$. We treat separately $\chi(z)$ as a 1D confinement  problem within a slab of thickness equal to that of the disk. This defines a TE effective index $n _{e}$ of the slab, and we obtain two separate Maxwell equations in cylindrical coordinates: $\frac { { { d }^{ 2 } } }{ d{ \theta }^{ 2 } } \Lambda (\theta )+{ m }^{ 2 }\Lambda (\theta )=0 $ and $ \left( \frac { { { d }^{ 2 } } }{ d{ r }^{ 2 } } +\frac { 1 }{ r } \frac { { { d } } }{ d{ r } }  \right) \Psi (r)+\left( \frac { { n }^{ 2 }_{e}{ \omega  }^{ 2 } }{ { c }^{ 2 } } -\frac { { m }^{ 2 } }{ { r }^{ 2 } }  \right) \Psi (r)=0 $ with m an azimuthal number. The solutions are $\Lambda (\theta )=A{ e }^{ -jm\theta }$ for the azimuthal function and Bessel and Hankel functions of the first and second kind for the radial function: 
\begin{equation}
{ \Psi }_{m}(r)=\begin{cases} { J }_{ m }(kn_{e}r)\qquad \text{for $r \le R$} \\ B{ H }_{ m }^{ (2) }(kr)\qquad \quad  \text{for $r>R$}  \end{cases}
\label{WGM_z_comp}
\end{equation}
with $R$ the disk radius and $B={ J }_{ m }(kn_{e}R){/ H }_{ m }^{ (2) }(kR)$. For each m, boundary conditions at the disk sidewall impose a discrete series of solutions $k_{m,p}$ labeled with a radial number $p$. For the dominant field component of the WGM, the azimuthal (radial) number $m$ ($p$) dictates the number of lobes within the disk in the azimuthal (radial) direction. In concordance with this circular classification of optical modes, there is a circular classification of excitonic modes as well, imposed by the cylindrical symmetry of the QW \cite{supp}. The exciton center-of-mass wave function is expressed in the QW plane in terms of Bessel functions of the first kind, this time with azimuthal (radial) number $m^{\prime}$ ($p^{\prime}$) : $F_{m^{\prime},p^{\prime}}(r,\theta)=K_{m^{\prime},p^{\prime}}J_{m}\left(\frac{x_{m^{\prime},p^{\prime}}r}{R}\right)e^{- j m^{\prime}\theta}$, where $K_{m^{\prime},p^{\prime}}$ is a normalization constant and $x_{m^{\prime},p^{\prime}}$ the $p^{\prime}$ root of the Bessel function of order $m^{\prime}$. In the supplements \cite{supp} we show that coupling occurs between optical and excitonic modes sharing the same azimuthal symmetry $m=m^{\prime}$ and that the Rabi splitting for a single QW positioned at the neutral plane of the disk takes the analytical form:

\begin{eqnarray}
\hbar \Omega^{m,p,p^{\prime}}_{R}= 2{\hbar}\sqrt{\frac{ e^{2}}{2\varepsilon_{0}\varepsilon_{r}m_{0}V_{\text{eff}}^{m,p}}\frac{f}{S}}C_{m,p,p^{\prime}}\nonumber\\
\int_{0}^{R}{drJ_{m}(\frac{x_{m,p^{\prime}}r}{R})J_{m}(k_{m,p}n_e r)}
\label{Rabi}
\end{eqnarray}

with $\varepsilon_{r}$ the relative permittivity of the semiconductor, $m_{0}$ the free electron mass, $V_{\text{eff}}^{m,p}$ the mode volume of the considered (m,p) WGM, $\frac{f}{S}$ the oscillator strength of the QW transition per unit surface and 
\begin{eqnarray}
C_{m,p,p{\prime}}=\frac{2\pi m}{[\max(J_{m-1}^2(k_{m,p}n_e r)+J_{m+1}^2(k_{m,p}n_e r))]^{1/2}}\nonumber\\
\frac{1}{R\sqrt{\pi}k_{m,p}n_e\vert J_{m-1}(x_{m,p^\prime}) \vert}
\end{eqnarray}

which we obtain by direct quantization of the fields when expressing the interaction part of the Hamiltonian $H_{int}= \sum_{m,p,p^{\prime}}\frac {\hbar} {2} \Omega^{m,p,p^{\prime}}_{R}\left(\hat{a}_{m,p}^{\dagger}\hat{d}_{m,p^{\prime}}+\hat{a}_{m,p}\hat{d}_{m,p^{\prime}}^{\dagger}\right)$, where the bosonic operators $\hat{a}_{m,p}$ and $\hat{d}_{m^{\prime},p^{\prime}}$ are respectively for photon and exciton modes. Equation \ref{Rabi} indicates that $p$ and $p^{\prime}$, while not necessarily equal, should be close to obtain maximal coupling. When $p=p^{\prime}$we obtain values of Rabi splitting between $6.6$ and $10.2$ meV, function of the considered WGM, and taking into account the presence of the five distributed QWs \cite{supp}. This is in very good agreement with values obtained in our experiments, which are of $6.64$, $5.79$ and $9.14$ meV for LP2, 3 and 5 respectively \cite{supp}.

\section{\label{sec:level5}Polariton lasing}
\label{lasing}

\begin{figure}
\centering
  \includegraphics[width=0.5\textwidth]{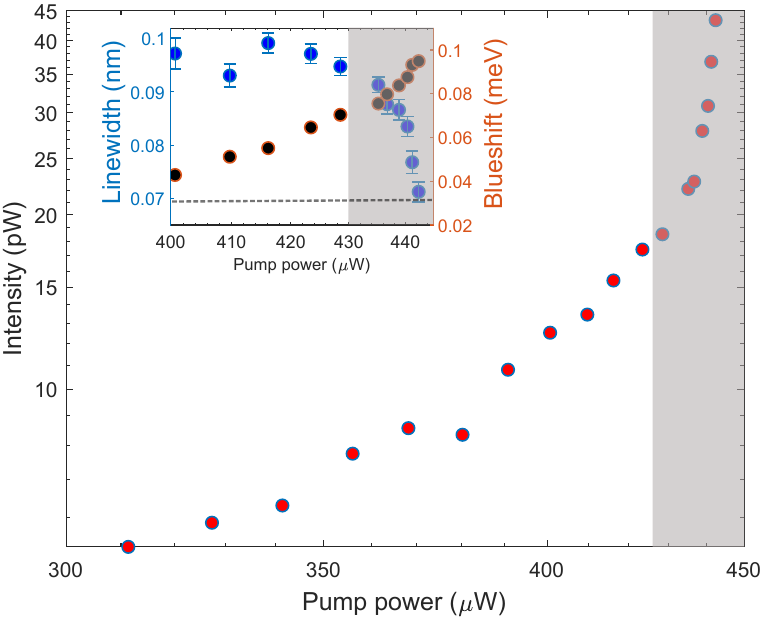}
    \caption{PL signal intensity (log-linear scale) as a function of the excitation power. Emitted power measured by the optical spectrum analyser analysing the emission (red circles, main plot). Emission blueshift (black circles, inset). Emission line width (blue circles, inset). The black dotted lines in the inset represent the spectral resolution of the optical spectrum analyser. The experiments are run under continuous wave optical pumping in configuration 2 (pump through the confocal path, and emission collection through the waveguide). }
\label{fig:5}
\end{figure}

Figure \ref{fig:5} shows the emission intensity of a single LP branch, as function of the excitation pump power $P$, for a continuous wave excitation at a wavelength of $840$ nm. The emission blueshift and line width are shown in the inset. Different regimes can be observed. For $P$ < 430 $\mu$W, as $P$ increases we observe a quasi-linear increase in the PL signal, with a relatively constant line width, while a moderate blue shift is already observable but remains much smaller than the Rabi energy. Above $430$ $\mu$W (gray shaded area in figure \ref{fig:5}), we observe a sharp nonlinear increase of the emission signal, together with a pronounced line narrowing, which is consistent with a lasing regime. At the same time, there is a marked inflexion of the blueshift. The latter remains however smaller than the Rabi splitting, consistent with the idea that the strong coupling regime persists above the observed lasing threshold. The combination of all these effects indicates that we are observing polariton lasing, sometimes dubbed non-equilibrium Bose-Einstein condensation of polaritons. This regime seems only to prevail for a certain range of excitation power, and in the supplemental document, we show data at even higher P, where the polariton lasing regime seems to get lost.

\section{\label{sec:level6}Conclusion}

In conclusion, we demonstrated the generation of quantum-well exciton polaritons in WGM InGaAs resonators. We observed the photoluminescence of numerous polariton branches, whose dispersion could be fitted in an unambiguous manner by a Hopfield model. All parameters of the model could be determined independently thanks to our original experimental configuration, allowing both non-resonant confocal microscopy and in-plane resonant optical spectroscopy. In addition, the Rabi coupling 
was computed using a new analytical model that we developed for circular polaritons formed between WGMs and QW excitons. In the non-linear regime, when the optical pumping was increased, we observed nonlinear increase of the emission, together with its line-narrowing and blueshift, which are all consistent with the regime of polariton lasing. The major hallmarks of the physics of polaritons are hence demonstrated in the present platform.

The present suspended GaAs disk resonators embedding InGaAs multiple quantum wells can serve as optomechanical resonators as well \cite{ding}, and it was recently predicted that enhanced optomechanical interactions could emerge in such case \cite{carlon2022enhanced}. Amongst other possibilities, WGM resonators will hence serve the purpose of exploring strong tripartite interactions between excitons, photons and mechanical motion. More generally, the here-developed tools and models for circular WGM polaritons should remain valid for many material platforms, notably those where the Q of WGMs reaches ultimate values. This creates the hope of stabilizing ultra-low dissipation polariton particles in the future, of interest in numerous domains of in light-matter interaction.

\begin{acknowledgments}
This work was supported by the European Research Council via the project the Consolidator grant NOMLI (770933). We thank Stephan Suffit and Pascal Filloux for advice and assistance in the cleanroom.
\end{acknowledgments}

\bibliography{apssamp,sample_SuppMod}

\newpage
\onecolumngrid

\title{Whispering gallery quantum well exciton polaritons in an Indium Gallium Arsenide microdisk cavity: supplements}%

\author{Romain de Oliveira }
\thanks{These authors contributed equally}
 \affiliation{Matériaux et Phénomènes Quantiques, Université Paris Cité, CNRS UMR 7162, 10 rue Alice Domon et Léonie Duquet 75013 Paris, France}
 
\author{Martin Colombano}
\thanks{These authors contributed equally}
\affiliation{Matériaux et Phénomènes Quantiques, Université Paris Cité, CNRS UMR 7162, 10 rue Alice Domon et Léonie Duquet 75013 Paris, France}

\author{Florent Malabat}
\affiliation{Matériaux et Phénomènes Quantiques, Université Paris Cité, CNRS UMR 7162, 10 rue Alice Domon et Léonie Duquet 75013 Paris, France}

\author{Martina Morassi}
\affiliation{Centre de Nanosciences et Nanotechnologies, CNRS UMR 9001, Université Paris-Saclay, 91120 Palaiseau, France}
\author{Aristide Lemaître}
\affiliation{Centre de Nanosciences et Nanotechnologies, CNRS UMR 9001, Université Paris-Saclay, 91120 Palaiseau, France}
\author{Ivan Favero}
\email{ivan.favero@u-paris.fr} 
\affiliation{Matériaux et Phénomènes Quantiques, Université Paris Cité, CNRS UMR 7162, 10 rue Alice Domon et Léonie Duquet 75013 Paris, France}

\section{Exciton and cavity energy temperature variation}

When the cryostat temperature increases, we observe a redshift in both the exciton and WGMs energy. (see Fig.2 in main text).  The temperature variations of the MQW excitonic energy are in agreement with predictions of three distinct models Varshni (Eq. \ref{eq:1}), Viña (Eq. \ref{eq:2}) and Passler (Eq. \ref{eq:3}) \cite{varshni,vina,passler} : 

\begin{align}
& E_{X}(T)=E_{X}(0)-\alpha\frac{T^{2}}{\beta+T} \label{eq:1}\\
& E_{X}(T)=E_{B}-a_{B}\left[1+\frac{2}{\exp(\Theta_{b}/T)-1}\right] \label{eq:2}\\
& E_{X}(T)=E_{X}(0)-\frac{\alpha\Theta}{2}\left[\sqrt[p]{1+\left(\frac{2T}{\Theta}\right)^{p}}-1\right] \label{eq:3}
\end{align}

Those models were originally proposed to compute the bandgap energy in bulk semiconductor materials, but they are also used for QW excitons \cite{rojas}. The model of Varshni is purely empirical: $E_{X}(0)$ is the energy of the exciton at 0 K, and $\alpha$ and $\beta$ are fitting parameters. Viña's model is a semi-empirical model based on the Bose-Einstein distribution:  $E_{B}-a_B$ is the energy at 0 K and $\Theta_{b}\equiv\hbar\omega/k_{B}$ is the effective phonon energy, expressed on the temperature scale. The last model, Passler's, is semi-emprical as well: it takes into account the effect of the lattice expansion and the electron-phonons interaction. $E_{X}(0)$ represents the energy of the exciton at 0 K, $\alpha$ is the high-temperature limit for the forbidden entropy, $\Theta$ the effective average phonon temperature and $p$ a parameter related to the shape of the electron-phonon spectral function. The fitting by these three models is represented in Figure \ref{fig:1}(a) and in the Table \ref{tab:1}. They all lead to a good agreement with experimental data. For the rest of our analysis we choose to fit the exciton energy with the model of Viña (main text), since it uses only 3 fitting parameters. \\

\begin{figure}[h!]
\centering
     \includegraphics[width=0.8\textwidth]{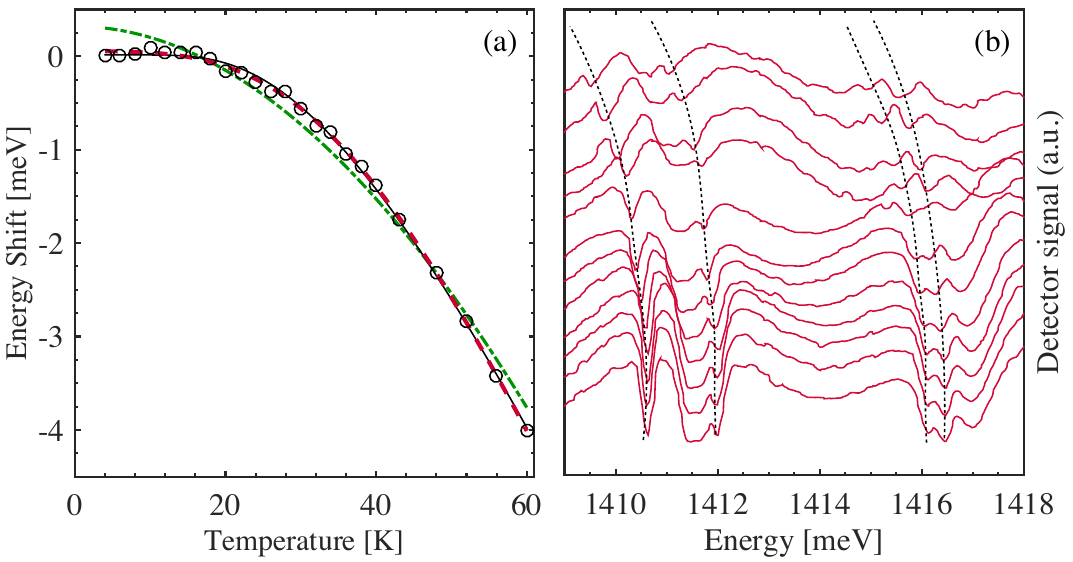}
   \caption{ (a) Fitting models for the temperature evolution of the exciton energy - Black open circles: Experimental data - Green dashed line: Varshni’s model ($r^2 \simeq 0.98$) - Red Solid line: Viña’s Model ($r^2 \simeq 0.99$) - Green dotted line: Passler’s Model ($r^2 \simeq 0.99$). (b) WGM spectra acquired in the waveguide transmission as function of the cryostat temperature. the black dotted lines are guides to the eye. }
\label{fig:1}
\end{figure}

\begin{table}[]
\begin{tabular}{cccccc}
\hline
Model   & \multicolumn{5}{c}{Parameter}                                      \\ \hline
Varshni & $E_X(0)$ [eV] & $\alpha$ [$\times 10^{-4}$ eV/K]& $\beta$[K]     & ... & $r^2$ \\ \hline
& $ 1.4732 \pm 4.41\times 10^{-6}$ & $11.999 \pm 2.2$                    & $998.9 \pm 195.5$  & ...& 0.981 \\ \hline
Viña & $E_B$ [eV]& $a_B$ [eV] & $\Theta_b$ [K] & ...& $r^2$ \\ \hline
& $1.4832 \pm 3.51\times 10^{-5}$  & $0.01033 \pm 3.57 \times 10^{-5}$   & $109.29\pm 0.165$  & ... & 0.998 \\ \hline
Passler & $E_X(0)$ [eV] & $\alpha$ [$\times 10^{-4}$ eV/K]& $\Theta$ [K] & $p$ & $r^2$ \\ \hline
& $ 1.4729 \pm 1.13\times 10^{-6}$ & $1.8354 \pm 0.012$  & $85.58\pm0.54$     & $3.442\pm 0.15$ & 0.999
\end{tabular}
\caption{Fitting parameters for the Varshni, Viña ans Passler models, $r^2$ represents the correlation coefficient.}
\label{tab:1}
\end{table}

The redshift of optical WGMs with temperature increase is mostly due to variation of the refractive index  (thermal expansion of the disk is negligible in comparison). Modeling these thermo-optic variations at cryogenic temperature is a tricky task, and works usually treat the refractive index with a Sellmeier equation having a varying number of poles \cite{wei,bertolotti,kisting,skauli,tanguy}. Unfortunately the parameters involved in such descriptions are usually measured far from our cryogenic temperatures (T$\simeq$ 3-4 K), hence we miss a simple model capable to describe the variation of GaAs refractive index in our range of temperatures. Working in a restricted wavelength range, we opted for using a mere Taylor expansion as function of temperature: 

\begin{equation}
    n={n_0+A_1 T+A_2 T^2 + A_3 T^3 + ...}
\end{equation}

where $n_0$ represents the refractive index at 0 K and $A_1$,$A_2$,$A_3$.. are fitting parameters to determine. Far from electronic transitions, we then employed an approximate relation for a WGM resonant wavelength : 

\begin{equation}
2\pi R n_{eff}\simeq m\lambda
\end{equation}

where $m$ represents the azimuthal number of the WGM, $R$ the radius of the microdisk and $n_{eff}$ the effective refractive index of the disk slab. Expressing this last condition in terms of energy and including the variations of the refractive index with temperature we obtain a custom expression to fit the variations of the WGMs energy : 

\begin{equation}
    E_{C}=\frac{mhc}{2\pi R}\frac{1}{n_{eff}^0+A_1 T+A_2 T^2 + A_3 T^3 + ...}
     \label{eq:0}
\end{equation}

The development at cubic power leads to a very good fit of experimental measurements, as shown in Figure 1 of main text. This expression was used to fit the variations of more than 20 distinct WGMs, and the fitting parameters $A_1$,$A_2$, and $A_3$ varied by less than 5 \%. The WGMs adjust to train this simple model were far in energy from any excitonic resonance,  in order to avoid influence of the MQW on temperature energy variations. In figure \ref{fig:1} (b) we present the waveguide transmission spectra as function of temperature, showing different WGMs resonances. The WGMs are separated by 20 to 30 meV from the MQW exciton, and hence behave as mere bare cavity modes (polaritonic effects can be neglected). \\ 

\section{Resonant waveguide spectroscopy of polaritons}

Figure \ref{fig:3} shows the LPs energy measured at 4K on a disk using the two distinct experimental configurations 2 and 3 introduced in figure 1 (c) of the main text: non-resonant confocal excitation and PL collection via the waveguides (Fig. \ref{fig:3} (a)) and resonant spectroscopy through the waveguide (Fig. \ref{fig:3} (b)). Except the fact that the energies are not measured by the same apparatus (hence slightly offset as result of calibration), their temperature variation, as well as the energy separation between the LP branches, seem to be consistent in both configurations. Both configurations have benefits and drawbacks. In resonant laser spectroscopy (configuration 3), an asset is to observe a larger number of LPs branches: indeed far-detuned LPs behave almost as photonic modes, and are consequently hardly visible in a PL spectrum. In contrast, they are resolvable in resonant laser spectroscopy through the waveguide. However, a drawback of this configuration is that the MQW being also present in the core of the waveguide, there is strong absorption as the laser light is tuned towards the exciton, compromising the resolution of close-by LP signals. Moreover because of this absorption, a good part of the injected light is lost in waveguide transmission experiment, and the estimation of the fraction of power injected into the WGMs is prone to errors. Finally, residual Fabry-Perot fringes in the waveguide transmission can make the collected signal more difficult to interpret (see Fig. \ref{fig:3} (b)). For these reasons, we often choose the experimental configuration 2, as it leads to a cleaner and stronger signal. \\

\begin{figure}[h!]
\centering
    \includegraphics[width=0.8\textwidth]{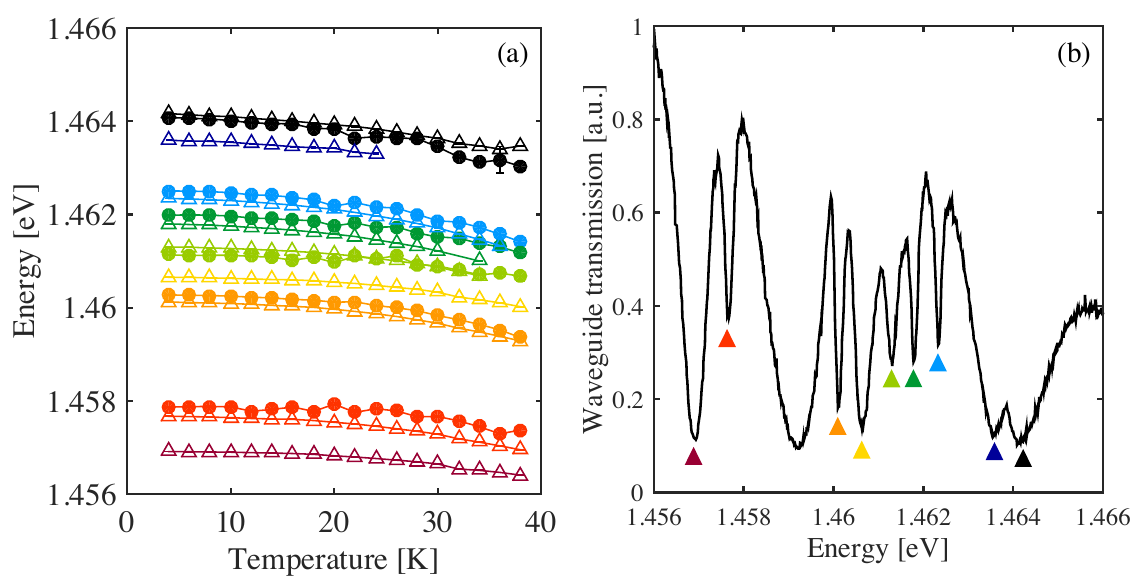}
    \caption{(a) Energy variation of 8 LP branches as function of temperature, for a disk of radius $R= 1.984$ $\mu$m. Circles: PL emission energy of LPs measured using configuration 2 of the set-up. Triangles: resonant laser spectroscopy measured using configuration 3 of the set-up. Black signal: exciton. Each colored branch corresponds to a distinct LP. The thin solid lines are guides to the eyes. (b) Waveguide optical transmission spectrum at 4 K for the same disk. The triangles highlight the exciton and LPs.}
\label{fig:3}
\end{figure}

\section{Polaritons spectrum fit by the Hopfield model}

We fit the temperature evolution of the eigenenergy of UPs and LPs with the Hopfield model expression: 

\begin{equation}
    E_{u/l}(T,m,p)= \frac{1}{2} \left( E_{x}(T)+E_{c}(T,m,p) \pm\hbar\sqrt{\delta^{2}(T,m,p)+\Omega_{R}^{2}}\right)
    \label{eq:7}
\end{equation}

where the $\pm$ sign stands respectively for UPs and LPs, and where  $\Omega_{R}$ is the Rabi splitting between the exciton and WGM ($m$,$p$). The temperature variation of the cavity energy ($E_c$) and exciton energy ($E_x$) have been discussed and modeled above and in the main text. As suggested by \eqref{eq:0}, we do not explicitly consider the role of the radial number $p$ in the temperature variation of \eqref{eq:7}. Finally, we are hence let with a fitting expression that contains three fitting parameters for each polariton branch of a given disk: $E_x(T=4K)$, $E_c(T=4K)$ (which is determined by $m$ according to \eqref{eq:0}), and the Rabi splitting $\Omega_{R}$, which we assume independent of temperature. Because $E_x$ is always measurable in the experiment, it is not an adjustable parameter in the fitting procedure and only two parameters need to be determined for a given branch: $\delta_{0}=\delta(4 K)$ (in practice $m$, according to \eqref{eq:0}) and $\Omega_{R}$. In order to determine these two parameters in a unique manner, in our experiments we measure disks whose radius varies by a few nanometers, which varies $\delta_{0}$ while letting $\Omega_{R}$ and $m$ constant. By fitting simultaneously the polariton spectra of these distinct disks, together with their temperature dispersion, we obtain for each LP branch a unique number $m$ and a unique value of $\Omega_{R}$. 

Figure \ref{fig:2} represents the temperature evolution of the polaritons energy for three different disks having nearby radius, and for 2 distinct polariton modes (LPs 2 and 5, see Fig. 3 of main text). A similar figure for the LP3 branch is presented in the main text. The data are fitted simultaneously on 5 distinct disks of distinct radius, but only 3 are shown here for clarity. We see the influence of the disk radius on the detuning: decreasing the radius increases the initial detuning (see Table \ref{tab:2}). Depending on the sign of the initial detuning $\delta_0$, an anticrossing may or may not be observed as temperature increases, as illustrated in Fig. \ref{fig:4}. 

\begin{figure}[h!]
\centering
     \includegraphics[width=0.85\textwidth]{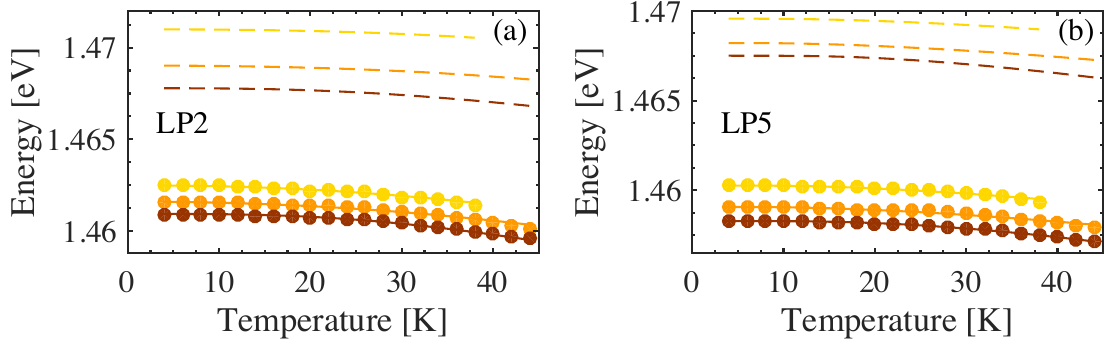}
    \caption{LPs energy as a function of temperature for 3 disks of distinct radius (Brown: $R$=1.992 $\mu$m - Orange: $R$=1.989 $\mu$m - Yellow: $R$= 1.984 $\mu$m). Circles: measured LPs emission energies. Solid: fits derived from the Hopfield model for the LPs. Dashed: UP lines derived from the Hopfield model - (a): LP2 - (b): LP5.}
\label{fig:2}
\end{figure}

\begin{figure}[h!]
\centering
     \includegraphics[width=0.85\textwidth]{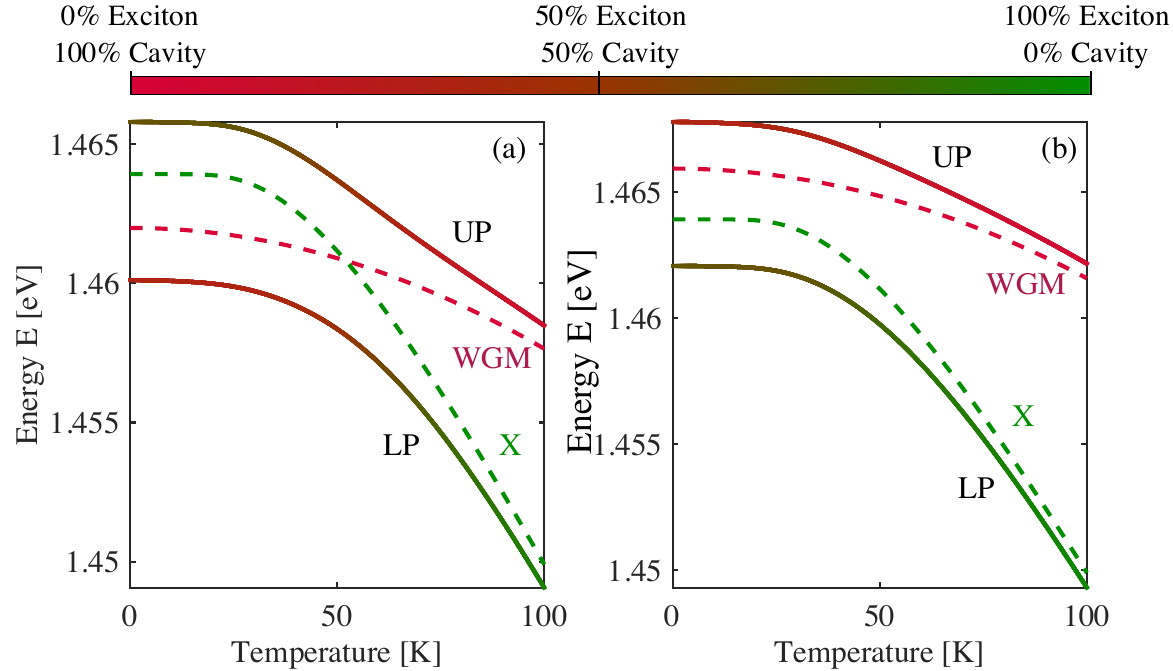}
    \caption{Energy dispersion of the exciton (X), the WGM and the UP/LP according to the Hopfield model (a) Negative initial detuning ($\delta_0 \simeq -2 \ \text{meV}$, ) (b) Positive initial detuning ($\delta_0 \simeq 2 \ \text{meV}$). UP and LP lines are color-coded according to the Hopfield coefficients. Green : exciton (X) - Red : WGM.}
\label{fig:4}
\end{figure}

\begin{table}[h]
\centering
\begin{tabular}{c c c c }
\hline
Radius R      & $\delta_0$ LP2 [meV] & $\delta_0$ LP3 [meV] & $\delta_0$ LP5 [meV] \\ \hline
1.992\ $\mu$m & 1.6897                 & - 0.9435                & - 1.2041                \\ \hline
1.989\ $\mu$m & 3.3248                & 0.2741              & 0.0390                \\ \hline
1.984\ $\mu$m & 5.3273                 & 2.0049                & 1.7335                \\ \hline
\end{tabular}
\caption{Evolution of the exciton-cavity mode initial detuning $\delta_0$, as function of the radius and for three distinct LP modes.}
\label{tab:2}
\end{table}

A potential improvement of the Hopfield fitting is to take into account the dissipation in order to express the complex eigenenergies: 

\begin{equation}
\begin{split}
     \tilde{E}_{u/l}(T,m,p)=& \frac{1}{2} \left[  \left(E_{x}(T)+E_{c}(T,m,p)\right)-i\hbar\left(\kappa_{c}(T,m,p)+\kappa_{x}(T)\right) \right] \\
     & \pm\frac{\hbar}{2} \left[ \sqrt{\left[\delta(T,m,p)-i(\kappa_{c}(T)-\kappa_{x}(T)\right]^{2}+\Omega_{R}^{2}(m,p)} \right]   
\end{split}
\end{equation}

where $\kappa_c$ ($\kappa_x$) is the dissipation rate for photons (excitons). Such dissipative fitting model requires the knowledge of temperature variations of these two rates. If the information on the exciton dissipation is directly measurable, it is less direct for the cavity modes. In our experimental configuration, the measured WGM line width includes information about the intrinsic losses of the WGM but also information about the extrinsic losses, i.e. the coupling of the WGM to the waveguide mode. $\kappa_c$ differs for each WGM, being in our experiments often comprised between 0.2 and 0.3 meV. Neglecting the contribution of the extrinsic losses leads to an underestimation of the Rabi splitting by 6\% for $\kappa_c$ = 0.3 meV, $\kappa_x$ = 2 meV and $\Omega_R$ = 5 meV. Note that the inhomogeneous broadening of the exciton is taken into account in this analysis. 

\section{Spectral signatures of polariton lasing}

In Fig. 5 of the main text, we showed clear signatures of polariton lasing observed on a LP branch with CW laser excitation at a wavelength of $840$ nm. In contrast, Fig. S5 groups the emission energy and intensity of all LP modes of a disk, as a function of the pump power $P$ measured in the confocal excitation path just before impinging on the disk. This enables collecting more signal and tracking the evolution over a larger range of excitation power. Here again different regimes are observed.
For $P$ < 100 $\mu$W, as the power increases we observe a slight increase in the PL signal, with relatively constant emission energies. For 100 $\mu$W < $P$ < 800 $\mu$W (gray shaded area in figure \ref{fig:5}), we observe a sharp nonlinear increase of the different PL signals, together with a marked blueshift on the energies, which is consistent with strong interactions. The amplitude of the blueshift remains smaller than Rabi splitting, consistent with the idea that the strong coupling regime persists above the threshold of this sharp nonlinear increase of the emitted intensity. For $P$ > 800 $\mu$W however, the signal intensity starts to decrease and the energies come back to their initial value. This may be due to an overpumping of the sample that destroys polaritons through heating or generation of an electron-hole plasma. In our present set of experiments, under continuous pumping, we observe an increase of the emitted signal by one order of magnitude before losing the lasing regime.

\begin{figure}[h!]
\centering
     \includegraphics[width=0.55\textwidth]{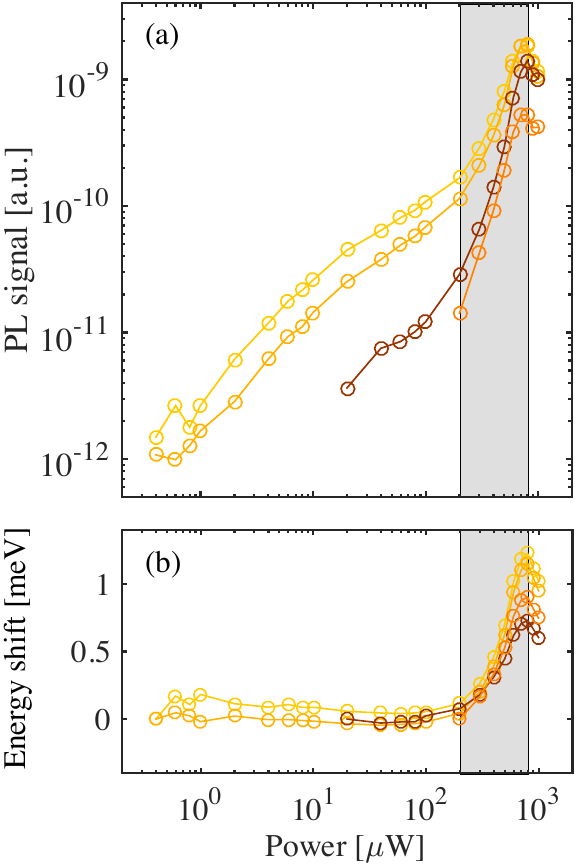}
    \caption{Supplementary data on polariton lasing. Polaritonic emission properties of four LP branches as function of the pump optical power impinging on the surface of the disk. The colors from brown to yellow indicate LP with an increasing energy. a) Emission intensity (log-log scale). b) Emission blue shift (linear-log scale).}
\label{fig:5}
\end{figure}

\section{Theoretical model for the exciton-WGM Rabi coupling}

\subsection{Electromagnetic whispering gallery modes of a dielectric disk}

The analysis starts with Maxwell equations without source terms, which leads to the Helmholtz equation in the dielectrics: 

\begin{equation}
\mathbf{\Delta}\mathbf{F}+\frac{n^{2}\omega^{2}}{c^{2}}\mathbf{F}=0
\label{waveeq}
\end{equation}

where $\mathbf{F}$ represents either the electric ($\mathbf{E}$) or magnetic field ($\mathbf{H}$) . The use of cylindrical coordinates $(r,\theta ,z)$ is natural in a disk. Inside our disk resonator, the thickness is smaller than the radius ($h\ll R$ in our system $h/R \simeq 0.1$) hence it seems reasonable to decouple the electromagnetic field variations along $z$ from those along $r$ and $\theta$, which helps transforming the dimensionality of the 3D problem into 2D+1. In the vertical z-direction, due to the step-index, the problem is treated as that of an infinite dielectric slab. The slab eigenwaves are either TE (transverse electric) or TM  (transverse magnetic), with an associated effective refractive index $n_e$. Once this vertical 1D problem is solved, the second step is to solve for the  propagation of electromagnetic in a flat (2D) disk of effective index $n_e$, considering this time a step-index in the radial direction. This 2D+1 approach is approximated, but it provides a useful analytic expression of the electromagnetic field. In this article we are mostly interested in TE modes, whose polarization enables coupling to the fundamental excitonic transition of the quantum wells. 

\subsubsection{Modes in a slab waveguide}
\label{slab_mode}

We consider an infinite slab waveguide and choose a mode propagating in the y-direction and invariant in the x-direction. The nonzero components of the electromagnetic field are written $\mathbf{F}(\mathbf{r}) = \mathbf{F}(x,z)e^{j(\omega t-\beta y)}$, with in the TE case $\mathbf{F}_{TE} = (E_x, H_y, H_z)$. Keeping the component in the plane of the slab and transverse to the propagation direction, the wave equation is:

\begin{equation}
\frac { { \partial  }^{ 2 } }{ { \partial z }^{ 2 } }F_{x}(z)+({ k }_{0}^{ 2 }{n(z)}^{2}-{ \beta}^{2 })F_{x}(z)=0
\label{eqa}
\end{equation}

with $k_{0}=\omega\sqrt{\mu_{0}\varepsilon_{0}}=\frac{2\pi}{\lambda}$ the propagation constant in vacuum and $\beta=k_{0}n_e$ the propagation constant along the y-axis. We consider a slab of thickness $h$ surrounded by vacuum hence $n(z)=n$ for $ |z|  \le \frac{h}{2} $ and $n(z)=1$ for $ |z|> \frac{h}{2}$. We are interested in the guided modes of the slab and use the ansatz of an oscillating field in the slab and exponentially decaying field in the surroundings:

\begin{align}
F_{x}(z)=A_{\sigma}{ e }^{ \gamma_{z}z}\quad &\text{for $z < -\frac{h}{2}$ }\\
F_{x}(z)=B_{\sigma}{ e }^{ jk_{z}z}+C_{\sigma}{ e }^{ -jk_{z}z}\quad & \text{for $ -\frac{h}{2} \le z  \le \frac{h}{2} $} \\ 
F_{x}(z)=D_{\sigma}{ e }^{ -\gamma_{z}z}\quad &\text{for $z > \frac{h}{2}$} 
\end{align}

with $k_z=\sqrt{n^{2}k_{0}^{2}-\beta^{2}}$ and $\gamma_z=\sqrt{\beta^{2}-k_{0}^{2}}$. $A_{\sigma},B_{\sigma},C_{\sigma}$ and $D_{\sigma}$ are constants with $\sigma$ the polarization TE or TM. The slab is symmetric, leading to identify two types of solutions: even and odd modes. In the case of vertically even modes the field solution is: 

\begin{align}
F_{x}(z)=A_{\sigma}{ e }^{ \gamma_{z}(z+\frac{h}{2})}\quad &\text{for $z < -\frac{h}{2}$ }\\
F_{x}(z)=B_{\sigma}\cos{k_{z}z}\quad & \text{for $ -\frac{h}{2} \le z  \le \frac{h}{2} $} \\ 
F_{x}(z)=A_{\sigma}{ e }^{ -\gamma_{z}(z-\frac{h}{2})}\quad &\text{for $z > \frac{h}{2}$} 
\end{align}

For odd modes, the field is given by: 

\begin{align}
F_{x}(z)=-A_{\sigma}{ e }^{ \gamma_{z}(z+\frac{h}{2})}\quad &\text{for $z < -\frac{h}{2}$ }\\
F_{x}(z)=B_{\sigma}\sin{k_{z}z}\quad & \text{for $ -\frac{h}{2} \le z  \le \frac{h}{2} $} \\ 
F_{x}(z)=A_{\sigma}{ e }^{ -\gamma_{z}(z-\frac{h}{2})}\quad &\text{for $z > \frac{h}{2}$} 
\end{align}

For TE modes, whose tangential component of the electric field (i.e. $E_{x}$) and tangential component of the magnetic field ($H_{y}$) must be continuous at $z=\pm\frac{h}{2}$, we obtain the following characteristic equations: 

\begin{align}
\frac{\gamma_z h}{2}& =\frac{k_z h}{2}\tan{\frac{k_z h}{2}} \quad \text{for even modes}
\label{eq:trans}
\\
-\frac{\gamma_z h}{2}& =\frac{k_z h}{2}\cot{\frac{k_z h}{2}} \quad \text{for odd modes}
\label{eq:trans_2}
\end{align}

Certain specific values of $\gamma_z$ and $k_z$ can satisfy these equations, i.e. the guide only supports a discrete set of modes. These equations cannot be solved analytically in a closed-form, and are treated with a numerical solver.

\subsubsection{Whispering gallery modes}
\label{WGMs_sec}

In the previous section, we obtained an effective index for the slab and calculated the dependence of the field in the z-direction. We consider now the effect of the radial confinement in the disk. In \eqref{eqa} the z-component $F_{z}$ can be treated independently from the two others. A separation of variables can be employed $F_{z}(r,\theta,z)=F_{z,\parallel}(r,\theta)F_{z,\perp}(z)=\Psi (r)\Lambda (\theta) X(z)$, where $F_{z,\parallel}$ is the field imposed by the radial confinement (in the plane of the disk) while $F_{z,\perp}$ is the field pattern imposed by the vertical confinement, calculated in the prior section \ref{slab_mode}. Inserting this ansatz in \eqref{eqa} leads to: 

\begin{equation}
\begin{cases} \frac { { { d }^{ 2 } } }{ d{ \theta }^{ 2 } } \Lambda (\theta )+{ m }^{ 2 }\Lambda (\theta )=0 \\ \frac { { { d }^{ 2 } } }{ d{ z }^{ 2 } } X(z)+\frac { { \omega  }^{ 2 } }{ { c }^{ 2 } } \left( { n }^{ 2 }-n_e^{ 2 } \right)X(z) =0 \\ \left( \frac { { { d }^{ 2 } } }{ d{ r }^{ 2 } } +\frac { 1 }{ r } \frac { { { d } } }{ d{ r } }  \right) \Psi (r)+\left( \frac { { n }^{ 2 }_{\text{e}}{ \omega  }^{ 2 } }{ { c }^{ 2 } } -\frac { { m }^{ 2 } }{ { r }^{ 2 } }  \right) \Psi (r)=0 \end{cases}
\label{em}
\end{equation}

The first equation is solved analytically $\Lambda (\theta )=A{ e }^{ -jm\theta  }$, where $m\in \mathbb{Z}$ is called the azimuthal number. The second equation is that of a slab waveguide solved previously. Introducing $u=(\omega/c)n_er$, the third equation reads: 

\begin{equation}
({ u }^{ 2 }\frac { { \partial  }^{ 2 } }{ \partial u^{ 2 } } +{ u }\frac { { \partial  } }{ { \partial u } } ){ \Psi }(r)  +(u^{ 2 }-{ m }^{ 2 }){ \Psi } (r)=0
\label{Bessel}
\end{equation}

If $u$ is a complex number, the resonance frequencies $\omega$ possess an imaginary part. The time dependence of solutions is $e^{j\omega t}$ and we consider $\omega =\omega_{0}+j\kappa$ with $\kappa > 0$ to grant convergence. The solutions of \eqref{Bessel} are Bessel and Hankel functions of the first and second kind ($J_{m}(u)$, $Y_{m}(u)$, $H_{m}^{(1)}(u)$,  $H_{m}^{(2)}(u)$). The field must be null at $r=0$ ($u=0$) inside the disk, and decrease radially towards zero outside the disk. The first condition is only verified by the Bessel function of first kind, and the second by the Hankel function of second kind. Hence: 

\begin{equation}
{ \Psi }(r)=\begin{cases} N{ J }_{ m }(k n_e r)\qquad \text{for $r \le R$} \\ NB{ H }_{ m }^{ (2) }(kr)\qquad \quad  \text{for $r>R$}  \end{cases}
\label{WGM_z_comp}
\end{equation}

with $n_e\simeq 1$ in vacuum ($r>R$). $N$ is a normalization constant and $B={ J }_{ m }(kn_eR){/ H }_{ m }^{ (2) }(kR)$. The tangential components of the electric and magnetic field must be continuous at the interface (${ H }_{ z } $, $ { E }_{ \theta  }  $ for TE polarization). We thus obtain the following characteristic equation: 

\begin{align}
\begin{split}
 \frac { { \dot { J }  }_{ m }(kn_eR) }{ { J }_{ m }(kn_eR) } =n_e\frac { { \dot { H }  }_{ m }^{ (2) }(kR) }{ { H }_{ m }^{ (2) }(kR) } 
\end{split}
\label{TETM} 
\end{align}

For a $m$ number, several solutions $k_{m,p}$ of equation \eqref{TETM} can exist: they are labeled with a radial number $p$. The azimuthal number $m$ represents the number of field oscillations around the disk, while its sign indicates if the phase propagation is ``clockwise" ($m>0$ ) or ``counter-clockwise" ($m<0$). The radial number $p$ corresponds to the number of lobes of the field in the radial direction. The other components of the field are obtained through (for $r<R$): 

\begin{equation}
 \begin{cases} { E }_{ r }=\frac { -m }{ { \omega \varepsilon _{ 0 }{ \varepsilon  }_{e}r } } { H }_{ z }==-\frac { N_{m,p}m }{ { \omega\varepsilon_{ 0 } { \varepsilon  }_{e}r } }{ J }_{ m }(\beta_{m,p}r)  e^{ -jm\theta } \\ { E }_{ \theta  }=\frac { j }{ { \omega { \varepsilon  }_{ 0 }{ \varepsilon  }_{e}  } } \frac { \partial { H }_{ z } }{ \partial r }=\frac { N_{m,p}j }{ { \omega \varepsilon_{ 0 }{ \varepsilon  }_{e} } } \left[ \frac { m }{ r }J_{m}(\beta_{m,p}r)-\beta_{m,p}J_{m+1}(\beta_{m,p}r) \right]e^{ -jm\theta }  \end{cases}
\end{equation}

with $\beta_{m,p}=k_{m,p}n_e$. We will later make use of a compact notation involving circularly polarized components of the electromagnetic field: the normalized complex vectors $\mathbf{E_{+}}$ and $\mathbf{E_{-}}$ defined by: 

\begin{equation}
\mathbf{E_{\pm}}=E_{\pm}\frac{(\mathbf{x}\mp j\mathbf{y})}{\sqrt{2}}e^{j\omega t}
\end{equation}

where $E_{\pm}=|E_{\pm}|e^{j\delta_{\pm}}$ is a complex-valued scalar quantity. Introducing the short-hand basis:  

\begin{equation}
\mathbf{v_{+}}=\frac{\mathbf{x}-j\mathbf{y}}{\sqrt{2}}=\frac{e^{-j\theta}}{\sqrt{2}}(\mathbf{r}-j\mathbf{\theta})=\mathbf{v_{-}^{*}} \quad \quad \quad \mathbf{v_{-}}=\frac{\mathbf{x}+j\mathbf{y}}{\sqrt{2}}=\frac{e^{j\theta}}{\sqrt{2}}(\mathbf{r}+j\mathbf{\theta})=\mathbf{v_{+}^{*}}
\label{eq:diff_base}
\end{equation}

one can write any electromagnetic field as a linear superposition: 

\begin{equation}
\mathbf{E} =E_{x}\mathbf{x}+E_{y}\mathbf{y}+E_{z}\mathbf{z} =E_{r}\mathbf{r}+E_{\theta}\mathbf{\theta}+E_{z}\mathbf{z}=E_{+}\mathbf{v_{+}}+E_{-}\mathbf{v_{-}}+E_{z}\mathbf{z}
\end{equation}

From the transfer matrices  

\begin{equation}
\begin{bmatrix}\mathbf{v_{+}}  \\ \mathbf{v_{-}}  \\ \mathbf{z} \end{bmatrix} =\frac{1}{\sqrt{2}}\begin{pmatrix} 1 & -j & 0 \\ 1 & j & 0 \\ 0 & 0 & 1 \end{pmatrix}\begin{bmatrix}\mathbf{x}  \\ \mathbf{y}  \\ \mathbf{z} \end{bmatrix} \quad , \quad
\begin{bmatrix}\mathbf{v_{+}}  \\ \mathbf{v_{-}}  \\ \mathbf{z} \end{bmatrix} =\frac{1}{\sqrt{2}}\begin{pmatrix} { e }^{ -j\theta  } & { -je }^{ -j\theta  } & 0 \\ { e }^{ j\theta  } & j{ e }^{ j\theta  } & 0 \\ 0 & 0 & 1 \end{pmatrix}\begin{bmatrix}\mathbf{r}  \\ \mathbf{\theta}  \\ \mathbf{z} \end{bmatrix} 
\end{equation}

we deduce the following relations: 

\begin{equation}
{ E }_{ \pm  } =\frac{({ E }_{ x }\pm j{ E }_{ y })}{\sqrt{2}}=e^{\pm j \theta}\frac{({ E }_{ r }\pm j{ E }_{ \theta })}{\sqrt{2}}
\end{equation}

For the TE WGMs we obtain (for $r<R$): 
 
\begin{align}
&E_{r}\pm jE_{\theta}=-\frac{N_{m,p}\beta_{m,p}}{\omega\varepsilon_{ 0 }\varepsilon_e}{ J }_{ m \mp 1 }(\beta_{m,p}r)e^{- jm\theta }\\
\end{align}

which gives the following expression for the circularly polarized electric field (for $r<R$):

\begin{equation}
\\mathbf{E_{\pm}}=-\frac{N_{m,p}\beta_{m,p}}{\sqrt{2}\omega\varepsilon_{ 0 }\varepsilon_e}{ J }_{ m\mp 1 }(\beta_{m,p}r)e^{-j(m\mp 1)\theta }\mathbf{v}_{\pm}
\label{Epm}
\end{equation}

The full 3D spatial distribution of the TE WGM can now be expressed, by multiplying the slab waveguide vertical variation (section \ref{slab_mode}) by the in-plane variations within the 2D disk (section \ref{WGMs_sec}). We get for $r<R$: 

\begin{align}
\begin{split}
\mathbf{E}_{m,p}^{\text{TE}}&=\left[\mathbf{E_{+}}(r,\theta)+\mathbf{E_{-}}(r,\theta)\right]E_{x,0}(z)=\mathbf{E}_{m,p}(r,\theta)E_{x,0}(z) \\
&=E_{0,m,p}^{\text{(TE)}}E_{x,0}(z)\left[{ J }_{ m-1 }(\beta_{m,p}r)e^{-j(m-1) \theta }\mathbf{v}_{+}+{ J }_{ m+1 }(\beta_{m,p}r)e^{-j(m+1) \theta }\mathbf{v}_{-}\right]
\label{eq:field_3D_TE}
\end{split}
\end{align}

where the normalization factors introduced in previous relations ($A_{\sigma}$,$B_{\sigma}$,$N^{\sigma}_{m,p}$...) have been merged into a single coefficient $E_{0,m,p}^{\sigma}$ with the index $0$ pointing to the considered fundamental mode of the slab. Using the expression \eqref{eq:field_3D_TE} the effective (3D) volume of the TE WGM takes the form: 

\begin{align}
\begin{split}
V_{\text{eff}} & =\frac{1}{M}\iiint{\left\langle \varepsilon(\mathbf{r}){\left| E(\mathbf{r})\right|}^{2}\right\rangle d\mathbf{V}},\quad M= \max\left \langle \varepsilon(\mathbf{r}){\left| E(\mathbf{r})\right|}^{2} \right \rangle \\
 & = \frac { 1 }{ 2M } \int_{-\infty}^{+\infty}\int_{0}^{2\pi}{\int_{ 0 }^{ +\infty }{ rdrd\theta dz \varepsilon_{ 0 }\varepsilon _{r}(r,z) \left[ E_{+}E_{+}^{*}(r,\theta) + E_{-}E_{-}^{*}(r,\theta)\right]{\left| E_{x,0}(z) \right|}^{2} } }	 \\				
\end{split}
\end{align} 

For a TE WGM with $m=41$ and $p=1$, the effective volume is equal to $0.181 \ {\mu m}^{3}\simeq 7.5 (\lambda/n)^{3}$ (i.e. $\simeq 7.2$\% of the microdisk volume) consistent with values reported in \cite{srinivasan,radulaski2015visible} for different wavelength range/material. FEM simulations predict a similar value $\simeq 0.179 \ {\mu m}^{3}$. \\

\subsubsection{Quantization of the electromagnetic field and vector potential}

This section aims at expressing the field quantum operators in terms of the normal modes found above, in order to treat the interaction between light and matter. Without following the strict procedure of quantization of the electromagnetic field using a Lagrangian approach, we map the Hamiltonian of the electromagnetic field onto that of an harmonic mechanical oscillator, identifying the corresponding momentum and position operators before expressing these in terms of annihilation and creation operators. We start back with Helmholtz equation \eqref{waveeq} and write a solution as an expansion over normal modes, where the time dependence and the spatial variations are separated:\\

\begin{equation}
\mathbf{E}(\mathbf{r},t)=\sum_{n}{A_{n}(t)\mathbf{u_n}(\mathbf{r})}
\end{equation} 
 
The normal modes $\left\{\mathbf{u_{n}} \right\} $ form a basis and obey the bulk and boundary conditions of Maxwell's equations:

\begin{equation}
\nabla^{2}u_{n}=-k_{n}^2u_{n} \qquad \nabla\cdot\mathbf{u_{n}}=0 \qquad \mathbf{n_{\perp}}\times \mathbf{u_{n}}=\\athbf{0}
\label{eq:e_exp}
\end{equation}

where $\mathbf{n_{\perp}}$ is a unit vector normal to the boundary surface \footnote{This third condition is imposed because the mode are considered transverse toward infinity}. The modes also satisfy the orthonormality condition: 

\begin{equation}
\int{\mathbf{u_n}(\mathbf{r})\mathbf{u_{n'}}^{*}(\mathbf{r})d^{3}\mathbf{r}}=\delta_{n,n'}
\end{equation}

Substituting the expression \eqref{eq:e_exp} in the wave equation leads to an equation for $A_{n}(t)$: 

\begin{equation}
\sum_{n}\frac{d^{2}A_n(t)}{dt^2}+\frac{c^2}{n^2}k_{n}^2A_{n}(t)=0
\end{equation} 

Since the modes are independent : 

\begin{equation}
\frac{d^{2}A_n(t)}{dt^2}+\frac{c^2}{n^2}k_{n}^2A_{n}(t)=0
\end{equation}

This equation is that of an harmonic oscillator with frequency $\omega_{n}=ck_{n}/n$, hence $A_n(t)\propto e^{\pm j\omega_n t}$. In a similar fashion we can express the magnetic field with a separation of variables, $\mathbf{H}(\mathbf{r},t)=\sum_{n}{B_{n}(t)\nabla\times\mathbf{u_n}(\mathbf{r})}$, where the coefficient $B_{n}$ must satisfy Maxwell's equations: 

\begin{align}
\begin{split}
\nabla \times \mathbf{E}=-{ \mu  }_{ 0 }\frac { \partial \mathbf{H} }{ \partial t } \ & \Rightarrow \ \sum _{ n }{ { A }_{ n }(t)\nabla \times  } \mathbf{ u_{ n }}=-{ \mu  }_{ 0 }\sum _{ n }{ { { \partial  }_{ t }B }_{ n }(t)\nabla \times  } \mathbf{u_{ n }} \\
& \Rightarrow  \frac { d{ B }_{ n }(t) }{ dt } = -\frac{1}{{ \mu  }_{ 0 }}{ A }_{ n }(t)
\end{split}
\end{align}

Using Maxwell-Ampere equation, $B_n$ obeys an harmonic oscillator equation as well: 

\begin{equation}
\frac{d^{2}B_n(t)}{dt^2}+\frac{c^2}{n^2}k_{n}^2B_{n}(t)=0
\end{equation}

For the sake of clarity, we write the energy as function of the real fields $\mathbf{E}_0$ and $\mathbf{H}_0$:

\begin{equation}
\hat{H} = \frac{1}{2}\int d^3\mathbf{r}\left(\varepsilon\mathbf{E}_0^2+\mu_{0}\mathbf{H}_0^{2}\right)
\end{equation}

and substitute the expression of the electric and magnetic field to obtain: 

\begin{equation}
\begin{split}
\hat{H} &=\frac{1}{2}\sum_{n',n}\left(\varepsilon A_{n}(t)A_{n'}(t)\int\mathbf{u_n}(\mathbf{r})\mathbf{u_{n'}}(\mathbf{r})d^3\mathbf{r}+ \right. \\
& \left. \qquad \qquad \mu_{0}B_{n}(t)B_{n'}(t)\int(\nabla\times\mathbf{u_n}(\mathbf{r}))\cdot(\nabla\times\mathbf{u_{n'}}(\mathbf{r}))d^3\mathbf{r}\right)\\
& = \sum_{n}\frac{1}{2}\left(\varepsilon A_{n}^2(t)+\mu_{0}k_{n}^{2}B_{n}^{2}(t)\right)
\end{split}
\end{equation}

where we have used $\int{(\nabla\times\mathbf{u_n}(\mathbf{r}))\cdot(\nabla\times\mathbf{u_{n'}}(\mathbf{r}))d^3\mathbf{r}}=k_{n}^{2}\delta_{n,n'}$. The electromagnetic Hamiltonian is hence similar to the Hamiltonian of a set of harmonic oscillators: 

\begin{equation}
\hat{H}_{h.o.}=\sum_{n}\frac{1}{2}\left(\frac{P_{n}^{2}(t)}{2m}+m\omega_{n}^{2}Q_{n}^{2}(t)\right)
\end{equation}

with $Q_{n}$ and $P_n=\frac{dQ_{n}}{dt}$ the position and momentum. We can hence identify $A_{n}$ and  $B_{n}$ to an equivalent position and momentum: 

\begin{align}
{ Q }_{ n }(t)\quad & \Longleftrightarrow \quad { A }_{ n }(t)=\sqrt { \frac { m{ \omega  }_{ n }^{ 2 } }{ \varepsilon  }  } { Q }_{ n }(t) \\
{ P }_{ n }(t)\quad & \Longleftrightarrow \quad { B }_{ n }(t)=\sqrt { \frac { 1 }{ { \mu  }_{ 0 }{ k }_{ n }^{ 2 }m }  } { P }_{ n }(t)
\end{align}

The position and momentum are themselves quantized and associated to an operator:

\begin{align}
{ \hat{Q} }_{ n }(t)& =\sqrt { \frac { \hbar  }{ 2m\omega_{n}  }  } \left( { { \hat { a }  } }_{ n }^{ \dagger   }(t)+{ \hat { a }  }_{ n }(t) \right) \\
{ \hat{P} }_{ n }(t)& =j\sqrt { \frac { \hbar m{ \omega  }_{ n } }{ 2 }  } \left( { { \hat { a }  } }_{ n }^{ \dagger   }(t)-{ \hat { a }  }_{ n }(t) \right) 
\end{align}

with ${ \hat { a }  }_{ n }$ and ${ \hat { a }  }_{ n }^{\dagger}$ the annihilation and creation operator of the quantum harmonic oscillator (bosonic ladder operators). By analogy an operator is associated to the normal mode coefficients $A_{n}$ and $B_{n}$ : 

\begin{align}
{ \hat{A} }_{ n }(t)& =\sqrt { \frac { \hbar { \omega  }_{ n } }{ 2\varepsilon  }  } \left( { { \hat { a }  } }_{ n }^{ \dagger  }(t)+{ \hat { a }  }_{ n }(t) \right) \\
\quad { \hat{B} }_{ n }(t) & =j\frac{c}{n}\sqrt { \frac { \hbar  }{ { 2\mu  }_{ 0 }{ \omega  }_{ n } }  } \left( { { \hat { a }  } }_{ n }^{ \dagger  }(t)-{ \hat { a }  }_{ n }(t) \right) 
\end{align}

The electric and magnetic fields are defined as the sum over these normal modes. We can hence write the field operators at any position and time as \footnote{The evolution of the operator $\hat{a}_n(t)$ and $\hat{a}_n^{\dagger}(t)$ derives from the Heisenberg equation of motion $\frac{d\hat{a}_n}{dt} = -\frac{j}{\hbar}[\hat{H},\hat{a}_n(t)]=-j\omega_n \hat{a}_n$, hence $\hat{a}_n(t)=\hat{a}_n(0)e^{-j\omega_n t}$. For convenience we wrote $\hat{a}_n(0)$ as $\hat{a}_n$.}:

\begin{align}
\hat{E}(\mathbf{r},t) & =\sum_{n}\sqrt{\frac{\hbar\omega_{n}}{2\varepsilon}}\left(\hat{a}_{n}\mathbf{u}_{n}(\mathbf{r})e^{-j\omega_nt}+h.c\right)\\
\hat{H}(\mathbf{r},t) & =-\sum_{n}j\frac{c}{n}\sqrt{\frac{\hbar}{2\mu_0\omega_{n}}}\left(\hat{a}_{n}\nabla\times \mathbf{u}_{n}(\mathbf{r})e^{-j\omega_nt}-h.c\right)
\end{align}

Given these expressions, the Hamiltonian of the system now reads: 

\begin{equation}
\hat{H}=\sum_{n}\frac{\hbar\omega_{n}}{2}\left(\hat{a}_{n}^{\dagger}\hat{a}_{n}+\hat{a}_{n}\hat{a}_{n}^{\dagger}\right)=\sum_{n}\hbar\omega_{n}\left(\hat{a}_{n}^{\dagger}\hat{a}_{n}+\frac{1}{2}\right)
\end{equation}

This derivation is general and valid for an electromagnetic system governed by Helmoltz equation and \eqref{eq:e_exp}. Going back to the WGMs of a dielectric resonator, we select a set of normal modes satisfying the conditions of orthonormality and the conditions \eqref{eq:e_exp}: 

 \begin{equation}
 \mathbf{u}_{m,p}(\mathbf{r})=\frac{\mathbf{E}^{\sigma}_{m,p}(\mathbf{r})\sqrt{\varepsilon(\mathbf{r})}}{\sqrt{V_{\text{eff}}^{m,p}}\sqrt{\max\left(\varepsilon(\mathbf{r}){\left| \mathbf{E}^{\sigma}_{m,p}(\mathbf{r})\right|}^{2}\right) }}=\frac{\tilde{E}_{m,p}(\mathbf{r})\mathbf{e}_\sigma}{\sqrt{V_{\text{eff}}^{m,p}}}=\frac{\tilde{E}_{x,0}(z)\tilde{E}_{m,p}(r,\theta)\mathbf{e}_\sigma}{\sqrt{V_{\text{eff}}^{m,p}}}
 \end{equation}
 
with $\mathbf{e}_\sigma$ the polarization vector of the mode. In the case of WGMs, the set of harmonic oscillators is no longer labeled by a unique quantum number $n$ but by a doublet of azimuthal and radial numbers $m$ and $p$ \footnote{We consider in this article disks that are thin enough to only support the fundamental slab mode, hence we will not explicitly involve a third quantization number to parametrize the vertical confinement}. We can hence express the electric field operator for a disk resonator as: 

\begin{equation}
\hat{\mathbf{E}}(\mathbf{r},t)=\sum_{m=-\infty}^{+\infty}\sum_{p=1}^{+\infty}\sqrt{\frac{\hbar\omega_{m,p}}{2\varepsilon V_{\text{eff}}^{m,p}}}\left(\hat{a}_{m,p}\tilde{E}_{m,p}(\mathbf{r})e^{-j\omega_{m,p}t}\mathbf{e}_\sigma+h.c.\right)
\end{equation}

To obtain an expression of the vector potential operator, we need to choose a gauge. We choose the Coulomb gauge, where the vector potential is transverse ($\nabla\cdot\mathbf{A}=0$) and in the absence of free charges, the scalar potential is null ($\Phi=0$). In that case, the relation between the electric field and the vector potential is: 

\begin{equation}
\mathbf{E}=-\frac{\partial\mathbf{A}}{\partial t}
\end{equation}

Implying that the electric field is also a transverse vector. The vector potential operator expressed in the WGMs basis is: 

\begin{align}
\hat{\mathbf{A}}(\mathbf{r},t) & =j\sum_{m=-\infty}^{+\infty}\sum_{p=1}^{+\infty}\sqrt{\frac{\hbar}{2\varepsilon \omega_{m,p} V_{\text{eff}}^{m,p}}}\left(\hat{a}_{m,p}\tilde{E}_{m,p}(\mathbf{r})e^{-j\omega_{m,p}t}\mathbf{e}_\sigma-h.c.\right) 
\label{eq:vect_pot}
\end{align}

One can freely change the origin of the time in \eqref{eq:vect_pot} and obtain the following expression for the vector potential: 

\begin{align}
\hat{\mathbf{A}}(\mathbf{r},t) & =\sum_{m=-\infty}^{+\infty}\sum_{p=1}^{+\infty}\sqrt{\frac{\hbar}{2\varepsilon \omega_{m,p} V_{\text{eff}}^{m,p}}}\left(\hat{a}_{m,p}\tilde{E}_{m,p}(\mathbf{r})e^{-j\omega_{m,p}t^\prime}\mathbf{e}_\sigma+h.c.\right) 
\label{eq:vect_pot_2}
\end{align}

\subsection{Exciton in a Quantum well}

 In this article we are interested in excitons of a quantum well (QW), more precisely a type-I QW, where electrons and holes are confined in the same layer. In a QW the effective mass equation of the exciton system is given \cite{sugawara,bastard}:

\begin{equation}
\left( -\frac{\hbar^{2}}{2m_{c}^{*}}\nabla_{\mathbf{r}_{c}}^{2} -\frac{\hbar^{2}}{2m_{v}^{*}}\nabla_{\mathbf{r}_{v}}^{2}-\frac{e^{2}}{\varepsilon \left|\mathbf{r}_{c}-\mathbf{r}_{v} \right|} +V_{c}(\mathbf{r}_{c})+ V_{v}(\mathbf{r}_{v})-\bar{E} \right)\Psi_{env}(\mathbf{r}_{c},\mathbf{r}_{v})=0
\label{hameh}
\end{equation}

where $V_{c/v}(\mathbf{r}_{c/v})$ represents the QW confinement potential energy for the electron in the conduction and valence band respectively, with $\Psi_{env}$ the exciton envelope function, and $\bar{E}=E-E_{gap}$ the difference between exciton and gap energy. We have to introduce a z-direction QW 1D potential for carriers, which reduces the dimensionality from 3D to 2D. We assume these z-direction confinement potentials to be strong enough, and the well layer to be sufficiently thin, to confine the exciton in the $x-y$ plane (narrow well hypothesis). Therefore the Coulomb potential only depends on the in-plane separation $\\mathbf{\rho}= \mathbf{\rho}_{c}-\mathbf{\rho}_{v}$. For such strong confinement, we can separate the $z$ and in-plane variations of the potential: 

\begin{align}
V_{c}(\mathbf{r_{c}})=V_{c,\parallel}(\mathbf{\rho}_{c}) + V_{c,\perp}(z_{c}) \qquad
V_{v}(\mathbf{r_{v}})=V_{v,\parallel}(\mathbf{\rho}_{v}) + V_{v,\perp}(z_{v})
\end{align}

This couple of equation implies that the Hamiltonian \eqref{hameh} is separable in $z$ and $\mathbf{\rho}$, and the exciton wave function
can be written as:

\begin{equation}
\Psi_{env}(\mathbf{r}_{c},\mathbf{r}_{v})=\Phi(\mathbf{\rho}_{c},\mathbf{\rho}_{v})\chi_c(z_{c})\chi_v(z_{v})
\end{equation}

$\chi_c(z_{c})$ and $\chi_v(z_{v})$ are the eigenfunctions for a particle in a 1D box with rectangular potential. We restrict the discussion to the lowest order wave function $\chi$ associated to the fundamental state of this 1D potential in $z$-direction. The remaining problem is to solve the in-plane Hamitonian: 

\begin{equation}
\left( -\frac{\hbar^{2}}{2m_{c}^{*}}\nabla_{\mathbf{\rho}_{c}}^{2} -\frac{\hbar^{2}}{2m_{v}^{*}}\nabla_{\mathbf{\rho}_{v}}^{2}-\frac{e^{2}}{\varepsilon \left|\mathbf{\rho} \right|} +V_{c,\parallel}(\mathbf{\rho}_{c})+ V_{v,\parallel}(\mathbf{\rho}_{v})-\bar{E}+E_{c,\perp}+E_{v,\perp} \right)\Phi(\mathbf{\rho}_{c},\mathbf{\rho}_{v})=0
\label{hameh-plane}
\end{equation}

where $E_{c/v,\perp}$ are the confinement energies of the conduction/valence band electron in the 1D potential along $z$. By introducing the center of mass coordinates, we can expand the confinement potential in power of $\mathbf{\rho}$: 

\begin{align}
& \mathbf{\rho}=\mathbf{\rho}_{c}-\mathbf{\rho}_{v} \\
& \mathbf{R}_{\parallel}=\frac{(m_{c}^{*}\mathbf{\rho}_{c}+m_{v}^{*}\mathbf{\rho}_{v})}{M}\\
\begin{split}
&V_{c/v,\parallel}(\mathbf{\rho}_{c/v})=V_{c/v,\parallel}(\mathbf{R}_{\parallel}\pm\frac{m_{v/c}^{*}}{M}\mathbf{\rho})=V_{c/v,\parallel}(\mathbf{R}_{\parallel})+\nabla V_{c/v,\parallel}(\mathbf{R}_{\parallel})\frac{m_{c/v}^{*}}{M}\mathbf{\rho}+O({\mathbf{\rho}}^{2})\\
&\simeq V_{c/v,\parallel}(\mathbf{R}_{\parallel})
\end{split}
\end{align}

The last approximation is justified by the fact that the spatial scale of the 2D confinement is much larger than the Bohr radius of the exciton. The Hamiltonian \eqref{hameh-plane} can then be separated in the relative and center-of-mass in-plane coordinates. The exciton envelope function takes the following form : 

\begin{equation}
\Psi_{env}(\mathbf{r}_{c},\mathbf{r}_{v})=\Phi(\mathbf{\rho})F(\mathbf{R}_{\parallel})\chi_c(z_{c})\chi_v(z_{v})
\label{wafun}
\end{equation}

$\bullet$ $\Phi(\mathbf{\rho})$ is the solution of the 2D-hydrogen-like problem: 

\begin{equation}
\left( -\frac{\hbar^{2}}{2\mu}\nabla_{\mathbf{\rho}}^{2} -\frac{e^{2}}{\varepsilon \left|\mathbf{\rho} \right|}\right)\Phi(\mathbf{\rho})=E_{b,n}^{2D}\Phi(\mathbf{\rho})
\label{rho_exc_2D}
\end{equation}

where $E_{b,n}^{2D}$ is the 2D binding energy, which relates to 3D through \cite{murayama}: 

\begin{equation}
E_{b,n}^{2D}=4E_{b,n}^{3D} \quad	a_{B,2D}=a_{B,3D}/2
\end{equation}

However, in real QW structures the exciton is not exactly two dimensional, and its binding energy lies between $E_{b,n}^{3D}$ and $4E_{b,n}^{3D}$. Here  as well we will restrict our discussion to the lowest order orbital $n=1$, with the associated wave function: 

\begin{equation}
\Phi_{1}^{2D}(\mathbf{r})=\sqrt{\frac{2}{\pi}}\frac{2}{a_{B,2D}}e^{-2\mathbf{r}/a_{B,2D}} \quad \quad \Phi_{1}^{2D}(\mathbf{K})=\frac{\sqrt{2\pi}a_{B,2D}}{(1+a_{B,2D}^2\mathbf{K}^2/4)^{3/2}}
\end{equation}

$\bullet$ $F(\mathbf{R}_{\parallel})$ is the solution of the problem:

\begin{equation}
\left[-\frac{\hbar^{2}}{2M}\nabla_{\mathbf{R}_{\parallel}}^{2}+V_{c,\parallel}(\mathbf{R}_{\parallel})+V_{v,\parallel}(\mathbf{R}_{\parallel})\right]F(\mathbf{R}_{\parallel})=E_{\parallel}F(\mathbf{R}_{\parallel})
\label{hamcir}
\end{equation}

In the simple case of a free exciton in a infinite 2D plane $V_{c/v,\parallel}(\mathbf{R}_{\parallel})=0$.\\

The eigenenergies of a bound 2D exciton state are consequently:

\begin{equation}
E=E_{g}+E_{c,\perp}+E_{v,\perp}+E_{b,n}^{2D}+E_{\parallel}
\label{eq:nrj_exc}
\end{equation}

We define the creation operator: 

\begin{equation}
\hat{d}_{\alpha}^{\dagger}=\sum_{\mathbf{k},\mathbf{k'}}O_{\alpha}(\mathbf{k},\mathbf{k'})\hat{c}_{c,\mathbf{k}}^{\dagger}\hat{c}_{v,\mathbf{k'}}
\label{ex_crea_2D}
\end{equation}

where $O_{\alpha}(\mathbf{k},\mathbf{k'})$ is a Fourier transform of the exciton wave function \eqref{wafun}. The wave vector will be later-on split in its in-plane and out of plane components  $\mathbf{k}=(\mathbf{k}_{\perp},\mathbf{k}_{\parallel})$. $\alpha$ is a set of numbers that label and characterize the state of relative and center of mass motion in the plane. $\hat{c}_{v,\mathbf{k'}}$ and $\hat{c}_{c,\mathbf{k}}^{\dagger}$ are the annihilation and creation operators for Bloch electrons in the valence and conduction band respectively. In the $\mathbf{r}$-representation, the wave function for the electron is: 

\begin{equation}
\langle{ \mathbf{r} }\vert { { \hat { c }  }_{ \alpha \mathbf{k} }^{ \dagger  } }\vert { 0 } \rangle =\langle { \mathbf{r} }|{ \alpha \mathbf{k} } \rangle =\frac { 1 }{ \sqrt { V }  } { e }^{-j\mathbf{kr} }{ u }_{ \alpha \mathbf{k} }(\mathbf{r}), \quad \alpha \equiv c/v
\label{Bloch}
\end{equation}

with $V$ the quantization volume and ${ u }_{ \alpha \mathbf{k} }(\mathbf{r})$ the periodic Bloch function. Excitons obey a bosonic behaviour if the exciton density is smaller than a saturation density $n_{sat} \simeq 1/(2\pi a_{B,2D}^{2})$ \cite{ciuti2003}.

\subsubsection{Exciton in a circularly patterned quantum well}

So far we considered free excitons in a 2D-plane (therefore $F(\mathbf{R}_{\parallel})=e^{j\mathbf{k}_{\parallel}\cdot \mathbf{R}_{\parallel}}$). In the disk geometry, the QW layer is patterned with a circular symmetry. The wave function obeys then the Schrödinger equation: 

\begin{equation}
\left[-\frac{\hbar^{2}}{2M}\nabla^{2}_{\mathbf{R}_{\parallel}}+V(\mathbf{R}_{\parallel})\right]F(\mathbf{R}_{\parallel})=E_{\parallel}F(\mathbf{R}_{\parallel})
\label{hamcirII}
\end{equation}

with $V(\mathbf{R}_{\parallel})$ a potential defined by: 

\begin{equation}
V(\mathbf{R}_{\parallel})=\begin{cases} 0 \qquad \  \text{for $\mathbf{R}_{\parallel}<R$} \\ \infty \qquad \text{for $\mathbf{R}_{\parallel}>R$}  \end{cases}
\end{equation}

where $R$ is the radius of the disk. The polar-coordinates system $(r,\theta)$ is again natural to treat the problem. Equation \eqref{hamcirII} becomes: 

\begin{equation}
-\frac{\hbar^{2}}{2M}\left[\frac{\partial^{2}}{\partial r^{2}}+\frac{1}{r}\frac{\partial}{\partial r}+\frac{1}{r^{2}}\frac{\partial^{2}}{\partial \theta^{2}}\right]F(r,\theta)+V(r,\theta)F(r,\theta)=E_{\parallel}F(r,\theta)
\label{hamcir2}
\end{equation}

Substituting $F(r,\theta)=L(r)N(\theta)$ in \eqref{hamcir2} yields the following angular equation: 

\begin{equation}
\frac{d^{2}N(\theta)}{d\theta^{2}}=-m'^2N(\theta)
\label{anglular}
\end{equation}
 
and the following radial equation inside the circular potential well: 
 
 \begin{equation}
-\frac{\hbar^{2}}{2M}\left[\frac{\partial^{2}}{\partial r^{2}}+\frac{1}{r}\frac{\partial}{\partial r}-\frac{m'^2}{r^{2}}\right]L(r)=E_{\parallel}L(r)
\label{radial}
 \end{equation}
 
The normalized solutions of \eqref{anglular} are given by: 
 
 \begin{equation}
N(\theta)=\frac{1}{\sqrt{2\pi}}e^{- jm'\theta} \quad m' \in \mathbb{Z}
 \end{equation}
 
Taking $k=\sqrt{2ME_{\parallel}}/\hbar$, equation \eqref{radial} is transformed into: 
 
 \begin{equation}
\left[\frac{\partial^{2}}{\partial r^{2}}+\frac{1}{r}\frac{\partial}{\partial r}+\left(k^{2}-\frac{m'^2}{r^{2}}\right)\right]L(r)=0
 \end{equation}
 
The solutions of a this equation are given by the Bessel functions of first kind $J_{m'}(kr)$. For an infinite potential barrier we adopt the boundary condition  $L(R)=0$ and the allowed values of k are those satisfying the equation $J_{m'}(kR) = 0$. If we call $x_{m',p^{\prime}}$ the $p^{\prime}$-{\text{th}} root of the Bessel function of order $m'$, the allowed values of $k$ are  $k_{m',p^{\prime}}=x_{m',p^{\prime}}/R$, and the eigenenergies are given by $E_{\parallel,m',p^{\prime}} = \hbar^{2}x_{m',p^{\prime}}^{2} /2MR^{2}$.\\

For $m'\neq 0$, there are two eigenstates corresponding each to the eigenenergy $E_{m',p^{\prime}}$: 

\begin{equation}
F_{m',p^{\prime}}(r,\theta)=K_{m',p^{\prime}}J_{m'}\left(\frac{x_{m',p^{\prime}}r}{R}\right)e^{- jm'\theta}=L(r)N(\theta)
\label{wave_ex}
\end{equation}

where $K_{m',p^{\prime}}$ is a normalization constant. For $m'=0$ there is only one eigenstate. All eigenstates are normalized: 

\begin{equation}
\iint{rdrd\theta\left|F(r,\theta)\right|^2}=\int_{0}^{R}{rdr\left|L(r)\right|^2}\int_{0}^{2\pi}{d\theta\left|N(\theta)\right|^2}=1
\end{equation}

The function $N(\theta)$ requires a factor $\frac{1}{\sqrt{2\pi}}$ to be normalized, the normalization constant $K_{m',p^{\prime}}$ thus obeys: 

\begin{equation}
\int_{0}^{R}{rdr\left|K_{m',p^{\prime}}J_{m'}\left(\frac{x_{m',p^{\prime}}r}{R}\right)\right|^2}=\frac{1}{2\pi}
\end{equation}

Since we have:

\begin{equation}
\int_{0}^{R}{rJ_{m'}^{2}(kr)dr}=\frac{R^{2}}{2}\left(J_{m'}^{2}(kR)-J_{m'-1}(kR)J_{m'+1}(kR)\right)
\end{equation}

Using $J_{m'-1}(k_{m',p^{\prime}}R)=-J_{m'+1}(k_{m',p^{\prime}}R)$ the normalization constant $K_{m',p^{\prime}}$ is given by: 

\begin{equation}
K_{m',p^{\prime}}=\frac{\sqrt{2}}{R}\frac{1}{\sqrt{2\pi}}\frac{1}{\sqrt{J_{m'-1}(x_{m',p^{\prime}})J_{m'+1}(x_{m',p^{\prime}})}}=\frac{1}{R\sqrt{\pi}}\frac{1}{\left|J_{m'-1}(x_{m',p^{\prime}})\right|}
\label{norm_wave_exc}
\end{equation}

\subsection{Optoelectronical (photon-exciton) coupling}
\label{sec:oe}

\subsubsection*{Dipolar interaction}

We present here a derivation of the Rabi splitting for a quantum-well exciton embedded in a disk WGM resonator. This rigorous derivation is one of the original theoretical outcomes of this article. The Hamiltonian of the system we consider is: 

\begin{equation}
\hat{H}=\hat{H}_{ex}+\hat{H}_{EM}+\hat{H}_{I}
\end{equation}

with $\hat{H}_{ex}/\hat{H}_{EM}$ the excitonic (electromagnetic) Hamiltonian, and where we introduce the following interaction Hamiltonian (minimal coupling Hamiltonian \cite{savona1,panzarini,bassani,meystre}): 

\begin{equation}
\hat{H}_{I}=-\frac{e}{2m}\sum_{n}{\hat{\mathbf{A}}(\hat{\mathbf{r}}_{n})\cdot\hat{\mathbf{p}}_{n}+\hat{\mathbf{p}}_{n} \cdot \hat{\mathbf{A}}(\hat{\mathbf{r}}_{n})}+\frac{e^{2}}{2m}\sum_{n}{{\left|\hat{\mathbf{A}}(\hat{\mathbf{r}}_{n})\right|}^{2}}
\end{equation}

where $m$ is the electron mass, $\hat{\mathbf{r}}_{n}$ ($\hat{\mathbf{p}}_{n}$) are the position (momentum) operator of the n-th QW electron, and the sum runs over all the electrons in the system \footnote{Note that in this section we place ourselves in the Schrödinger ($S$) picture, where operators do not have an explicit time dependence. The expression of the vector potential $\hat{\mathbf{A}}$ differs from the one introduced in the Heisenberg picture ($H$) in equation \eqref{eq:vect_pot}. The relation between the two expressions is given by: $\hat{\mathbf{A}}^H=\hat{U}^\dagger\hat{\mathbf{A}}^S\hat{U}$ with $\hat{U}=e^{-j\hat{H}_{EM}t/\hbar}$  the evolution operator}. We leave aside the self-interaction term in ${\left|\hat{\mathbf{A}}(\hat{\mathbf{r}}_{n})\right|}^{2}$ and  focus on the exciton-photon interaction. In the Coulomb gauge the vector potential commutes with the electron momentum $\left[\hat{\mathbf{A}}(\hat{\mathbf{r}}_{n}),\hat{\mathbf{p}}_{n}\right] = 0$ since $\nabla \hat{\mathbf{A}} = 0$. The interaction term can be rewritten as: 

\begin{equation}
\hat{H}_{I}=\frac{je}{\hbar}\sum_{n}{\hat{\mathbf{A}}(\hat{\mathbf{r}}_{n})\cdot\left[\hat{\mathbf{r}}_{n},\hat{H}_{ex}\right]}
\label{hamint}
\end{equation}

since $\hat{\mathbf{p}}_{n}=j\frac{m}{\hbar}\left[\hat{H}_{ex},\hat{\mathbf{r}}_{n}\right]$ \footnote{This relation, formulated for single-particle operators, remains valid with Coulomb interactions as considered here. This is a consequence of the fact that $\hat{V}_{\text{Coulomb}}(r)$ commutes with $\hat{r}$ for the single considered particle}. The states $\left|\Psi_{\alpha}\right> = \hat{d}_{\alpha}^{\dagger} \left|\Psi_{0}\right>$ form a complete basis to describe the single exciton formed in the many-body system. We express the interaction Hamiltonian in terms of exciton wave-functions by inserting the unity operator twice and using the completeness relation: 

\begin{align}
&\sum_{\alpha}\left|\Psi_{\alpha}\right>\left<\Psi_{\alpha}\right|=\mathbb{I} \\
&\hat{H}_{I}=\frac{je}{\hbar}\sum_{\alpha,\beta}{\left|\Psi_{\alpha}\right> \left<\Psi_{\alpha}\right| {\sum_{n} {\hat{\mathbf{A}}(\hat{\mathbf{r}}_{n})\cdot\left[\hat{\mathbf{r}}_{n},\hat{H}_{ex}\right]\left|\Psi_{\beta}\right>}\left<\Psi_{\beta}\right|} } 
\end{align}

Using  $\hat{H}_{ex}\left|\Psi_{\alpha}\right>=E_{\alpha}\left|\Psi_{\alpha}\right>$, and the fact that $\hat{\mathbf{A}}(\hat{\mathbf{r}})$ commutes with $\hat{H}_{ex}$ in the Coulomb gauge, the equation can be rewritten as \footnote{$\left<\Psi_{\alpha}\right|\sum_{n}\hat{\mathbf{A}}(\mathbf{r}_{n})\cdot\mathbf{r}_{n}\left|\Psi_{\beta}\right>$ is an operator that only acts on the photon Hilbert space}: 

\begin{equation}
\hat{H}_{I}=\frac{je}{\hbar}\sum_{\alpha,\beta}(E_{\beta}-E_{\alpha})\left<\Psi_{\alpha}\right|\sum_{n}{\hat{\mathbf{A}}(\hat{\mathbf{r}}_{n})\cdot\hat{\mathbf{r}}_{n}\left|\Psi_{\beta}\right>\left|\Psi_{\alpha}\right>\left<\Psi_{\beta}\right|}
\label{ham_int1}
\end{equation}

\subsubsection*{Circular disk case}

We limit the exciton basis to the two lowest states, i.e. the ground state $\left|\Psi_{0}\right>$ and the first excited state $\left|\Psi_{m^{\prime},p^{\prime}}\right>$. We drop the index $\alpha/\beta$ and use the azimuthal/radial number $m^{\prime}/p^{\prime}$ instead, since our exciton state is fully characterized by those two quantum numbers. With those notations the interaction Hamiltonian \eqref{ham_int1} becomes: 

\begin{equation}
\begin{alignedat}{2}
& \hat{H}_{I} =je\sum_{m^{\prime}=-\infty}^{+\infty}\sum_{p^{\prime}=1}^{+\infty}\omega_{m^{\prime},p^{\prime}} && \left[-\left<\Psi_{m^{\prime},p^{\prime}}\right|\sum_{n}\hat{\mathbf{A}}(\hat{\mathbf{r}}_{n})\cdot\hat{\mathbf{r}}_{n}\left|\Psi_{0}\right>\left|\Psi_{m^{\prime},p^{\prime}}\right>\left<\Psi_{0}\right|\right]  \\
 & &&+ \left[  \left<\Psi_{0}\right|\sum_{n}\hat{\mathbf{A}}(\hat{\mathbf{r}}_{n})\cdot\hat{\mathbf{r}}_{n}\left|\Psi_{m^{\prime},p^{\prime}}\right>\left|\Psi_{0}\right>\left<\Psi_{m^{\prime},p^{\prime}}\right|\right] \\
& \hat{H}_{I} = je\sum_{m^{\prime}=-\infty}^{+\infty}\sum_{p^{\prime}=1}^{+\infty}\omega_{m^{\prime},p^{\prime}} && \left[ -\left<\Psi_{m^{\prime},p^{\prime}}\right|\sum_{n}{\hat{\mathbf{A}}(\hat{\mathbf{r}}_{n})\cdot\hat{\mathbf{r}}_{n}}\left|\Psi_{0}\right>\hat{d}_{m^{\prime},p^{\prime}}^{\dagger} \right] \\
&  &&+ \left[\left<\Psi_{0}\right|\sum_{n}{\hat{\mathbf{A}}(\hat{\mathbf{r}}_{n})\cdot\hat{\mathbf{r}}_{n}}\left|\Psi_{m^{\prime},p^{\prime}}\right>\hat{d}_{m^{\prime},p^{\prime}} \right] \\
& \hat{H}_{I} = e\sum_{m^{\prime}=-\infty}^{+\infty}\sum_{p^{\prime}=1}^{+\infty}\omega_{m^{\prime},p^{\prime}} && \left[ j\left<\Psi_{0}\right|\sum_{n}{\hat{\mathbf{A}}(\hat{\mathbf{r}}_{n})\cdot\hat{\mathbf{r}}_{n}}\left|\Psi_{m^{\prime},p^{\prime}}\right>\hat{d}_{m^{\prime},p^{\prime}} + h.c.\right] 
\end{alignedat}
\label{ham_int2}
\end{equation}

Where we identified $\left|\Psi_{m^{\prime},p^{\prime}}\right>\left<\Psi_{0}\right|$ as the creation operator $\hat{d}_{m^{\prime},p^{\prime}}^{\dagger}$ and $\left|\Psi_{0}\right>\left<\Psi_{m^{\prime},p^{\prime}}\right|$ as the annihilation operator $\hat{d}_{m^{\prime},p^{\prime}}$. We evaluate the expression $\hat{X}_{m^{\prime},p^{\prime}}=\left<\Psi_{0}\right|\sum_{n}{\hat{\mathbf{A}}(\hat{\mathbf{r}}_{n})\cdot\hat{\mathbf{r}}_{n}}\left|\Psi_{m^{\prime},p^{\prime}}\right> $ by expressing the action of $\sum_{n}{\hat{\mathbf{A}}(\hat{\mathbf{r}}_{n})\cdot\hat{\mathbf{r}}_{n}}$ on the many-electron states in the Bloch function basis. \\

According to the second quantization formalism \cite{cohen}, the expansion is given by:

\begin{equation}
\sum_{n}{\hat{\mathbf{A}}(\hat{\mathbf{r}}_{n})\cdot\hat{\mathbf{r}}_{n}}=\sum_{\mathbf{k},\mathbf{k}^{\prime},\alpha,\alpha^{\prime}}f_{\mathbf{k},\mathbf{k}^{\prime},\alpha,\alpha^{\prime}}\hat{c}_{\alpha,\mathbf{k}}^{\dagger}\hat{c}_{\alpha^{\prime},\mathbf{k}^{\prime}}
\end{equation}
 
where $\hat{c}_{\alpha,\mathbf{k}}^{\dagger}$,$\hat{c}_{\alpha^{\prime},\mathbf{k}^{\prime}}$ are the creation and annihilation operators introduced in \eqref{Bloch} and $f_{\mathbf{k},\mathbf{k}^{\prime},\alpha,\alpha^{\prime}}$ represents the matrix element: 



\begin{equation}
f_{\mathbf{k},\mathbf{k}^{\prime}\alpha,\alpha^{\prime}}=\left<\alpha\mathbf{k}\right|{\hat{\mathbf{A}}(\hat{\mathbf{r}})\cdot\hat{\mathbf{r}}}\left|\alpha^{\prime}\mathbf{k}^{\prime}\right>
\end{equation}


Using the orthogonality of the state $\hat{c}_{\alpha,\mathbf{k}}^{\dagger}\hat{c}_{\alpha^{\prime},\mathbf{k}^{\prime}}\left|\Psi_{0}\right>$ in association with the relation \eqref{ex_crea_2D} we obtain a new expression for the factor $X_{m^{\prime},p^{\prime}}$: 

\begin{equation}
\hat{X}_{m^{\prime},p^{\prime}}=\sum_{\mathbf{k},\mathbf{k}^{\prime}}f_{\mathbf{k},\mathbf{k}^{\prime},v,c}O_{m^{\prime},p^{\prime}}(\mathbf{k}^{\prime},\mathbf{k})
\end{equation}

We will compute $\hat{X}_{m^{\prime},p^{\prime}}$, knowing that its conjugate can be obtained through \footnote{$f_{\mathbf{k},\mathbf{k}^{\prime},v,c}^{*}=f_{\mathbf{k}^{\prime},\mathbf{k},c,v}$ since $\hat{\mathbf{A}}$ commutes with $\hat{\mathbf{r}}$ in the Coulomb gauge and both operators are hermitian.} $$\sum_{\mathbf{k},\mathbf{k}^{\prime}}f_{\mathbf{k},\mathbf{k}^{\prime},v,c}O_{m^{\prime},p^{\prime}}(\mathbf{k}^{\prime},\mathbf{k})=\left(\sum_{\mathbf{k},\mathbf{k}^{\prime}}f_{\mathbf{k}^{\prime},\mathbf{k},c,v}O_{m^{\prime},p^{\prime}}^{*}(\mathbf{k}^{\prime},\mathbf{k})\right)^{\dagger}$$

We employ an explicit expression for the matrix element $f$ for a TE optical mode \footnote{ Here $\tilde{E}_{m,p}^{\text{TE}}(\mathbf{r})\mathbf{e}_{TE}=\tilde{E}_{m,p}(\mathbf{r})\mathbf{e}(r,\theta)$, in the following the later vector is noted $\mathbf{e}$ for compactness}: 

\begin{align}
\begin{split}
f_{\mathbf{k},\mathbf{k}^{\prime},v,c}& =v\mathbf{k}{\hat{\mathbf{A}}(\hat{\mathbf{r}})\cdot\hat{\mathbf{r}}}{c\mathbf{k}^{\prime}}\\
&=  \sum_{m=-\infty}^{+\infty}\sum_{p=1}^{\infty}\frac{1}{V}\int d\mathbf{r}e^{-j(\mathbf{k}^{\prime}-\mathbf{k})\mathbf{r}}C_{m,p}\left[\hat{a}_{m,p}\tilde{E}_{m,p}(\mathbf{r})\mathbf{e}+\hat{a}_{m,p}^{\dagger}\tilde{E}^{*}_{m,p}(\mathbf{r})\mathbf{e}^*\right]\\
& \qquad \qquad \qquad \qquad \qquad \qquad \qquad \qquad \qquad \qquad \cdot u_{\mathbf{k}v}^{*}(\mathbf{r})\mathbf{r}u_{\mathbf{k}^{\prime}c}(\mathbf{r})
\end{split}
\label{matrix_el}
\end{align}

where $C_{m,p}=\sqrt{\frac{\hbar}{2\varepsilon_{0}n_{r}^{2}\omega_{m,p}V_{\text{eff}}^{m,p}}} $ and where V is the quantization volume for the Bloch functions. \\

The first term of equation \eqref{matrix_el} will give in the final Hamiltonian a term proportional to $\hat{d}_{m^{\prime},p^{\prime}}\hat{a}_{m,p}$, which we will be eliminated due to its non resonant nature. We therefore neglect this term, and reintroduce the matrix element $f$ into $\hat{X}_{m^{\prime},p^{\prime}}$ to obtain: 

\begin{align}
\begin{split}
\hat{X}_{m^{\prime},p^{\prime}} & =\sum_{m=-\infty}^{+\infty}\sum_{p=1}^{+\infty}C_{m,p}\hat{a}_{m,p}^{\dagger}\int d\mathbf{r}\left\lbrace \frac{1}{V}\sum_{\mathbf{k},\mathbf{k}^{\prime}}O_{m^{\prime},p^{\prime}}(\mathbf{k}^{\prime},\mathbf{k})  e^{-j(\mathbf{k}^{\prime}-\mathbf{k})\mathbf{r}} \right \rbrace  \tilde{E}^{*}_{m,p}(\mathbf{r})\mathbf{e}^*   \\
& \qquad \qquad \qquad \qquad \qquad \qquad \qquad \qquad \qquad \qquad \cdot u_{\mathbf{k}v}^{*}(\mathbf{r})\mathbf{r}u_{\mathbf{k}^{\prime}c}(\mathbf{r})
\end{split}
\label{int_before}
\end{align}

We place ourselves at the band edge, i.e. $\mathbf{k},\mathbf{k'}\simeq 0$, thus $u_{\mathbf{k}c/v}\simeq u_{0c/v}\equiv u_{c/v}$ and the only k-dependence is inside the curly braces \footnote{This approximation is justified since the $\mathbf{k}\cdot\mathbf{p}$ method that we will use later to determine momentum matrix element works for low value of $\mathbf{k}$.}. With this approximation, we identify the expression between curly braces as the Fourier transform of the exciton envelope function, taken with the electron and the hole at the same position $\mathbf{r}$: 

\begin{equation}
\frac{1}{V}\sum_{\mathbf{k},\mathbf{k}^{\prime}}O_{m^{\prime},p^{\prime}}(\mathbf{k}^{\prime},\mathbf{k})e^{-j(\mathbf{k}^{\prime}-\mathbf{k})\mathbf{r}}=\Phi(0)F_{m^{\prime},p^{\prime}}(\mathbf{R}_{\parallel})\chi_c(z_{c})\chi_v(z_{v})
\end{equation}

To evaluate the remaining integral \footnote{With electron and hole occupying the same position we have $z_c=z_v$ hence $\mathbf{r}=\mathbf{R}_{\parallel}+\mathbf{z}$}, 

\begin{equation}
I=\int d\mathbf{r} \Phi(0)F_{m^{\prime},p^{\prime}}(\mathbf{R}_{\parallel})\chi_c(z)\chi_v(z)\tilde{E}^{*}_{m,p}(\mathbf{r})\mathbf{e}^*\cdot u_{v}^{*}(\mathbf{r})\mathbf{r}u_{c}(\mathbf{r})
\label{int_before_2}
\end{equation}

we decompose the space into a sum of atomic unit cells: 

\begin{equation}
\int d\mathbf{r} \Longrightarrow \sum_{i}\int_{cell}d\mathbf{r}_{i}
\end{equation}

where $\mathbf{r}_{i} = \mathbf{r} - \mathbf{r}^{0}_{i}$ and $\mathbf{r}^{0}_{i}$ denotes the position of the i-th atom in the crystal. At the scale of a unit cell ($\simeq 0.1 $ nm), the functions appearing in the integral of \eqref{int_before_2} do not change noticeably, except for the atomic Bloch parts $u_{c}$ and $u_{v}^{*}$.

\begin{equation}
\begin{split}
I& =\sum_{i}\int_{cell}d\mathbf{r}_{i}\Phi(0)F_{m^{\prime},p^{\prime}}(\mathbf{R}_{\parallel,i}^{0}+\mathbf{R}_{\parallel,i})\chi_c(z_{i}^{0}+z_{i})\chi_v(z_{i}^{0}+z_{i})\tilde{E}^{*}_{m,p}(\mathbf{r}_{i}+\mathbf{r}_{i}^{0})\mathbf{e}^*\\
& \cdot u_{v}^{*}(\mathbf{r}_{i}+\mathbf{r}_{i}^{0})(\mathbf{r}_{i}+\mathbf{r}_{i}^{0})u_{c}(\mathbf{r}_{i}+\mathbf{r}_{i}^{0})\\
& \simeq \sum_{i}\Phi(0)F_{m^{\prime},p^{\prime}}(\mathbf{R}_{\parallel,i}^{0})\chi_c(z_{i}^{0})\chi_v(z_{i}^{0})\tilde{E}^{*}_{m,p}(\mathbf{r}_{i}^{0})\mathbf{e}^*\cdot \int_{cell}d\mathbf{r}_{i} u_{v}^{*}(\mathbf{r}_{i})\mathbf{r}_{i}u_{c}(\mathbf{r}_{i})
\end{split}
\end{equation}

The remaining integral is independent of $i$, and is precisely (up to the charge) the usual dipole matrix element between the valence and conduction-band atomic Bloch functions: 

\begin{equation}
\mathbf{r}_{vc}=\frac{1}{V_{cell}}\int_{cell}d\mathbf{r} u_{v}^{*}(\mathbf{r})\mathbf{r}u_{c}(\mathbf{r})=\langle u_v\vert\mathbf{r}\vert u_c\rangle
\end{equation}
 
By re-transforming the sum into an integral $\sum_{i} \Longrightarrow \frac{1}{V_{cell}}\int $ we obtain:

\begin{equation}
I = \mathbf{r}_{vc}\cdot \int{d\mathbf{r}\Phi(0)F_{m^{\prime},p^{\prime}}(\mathbf{R}_{\parallel})\chi_c(z)\chi_v(z)\tilde{E}^{*}_{m,p}}(\mathbf{r})\mathbf{e}^*
\end{equation}

and $\hat{X}_{m^{\prime},p^{\prime}}$ takes the following form: 

\begin{equation}
\hat{X}_{m^{\prime},p^{\prime}}=-j\sum_{m=-\infty}^{+\infty}\sum_{p=1}^{+\infty}C_{m,p}\hat{a}_{m,p}^{\dagger}\mathbf{r}_{vc}\cdot \int{d\mathbf{r}\Phi(0)F_{m^{\prime},p^{\prime}}(\mathbf{R}_{\parallel})\chi_c(z)\chi_v(z)\tilde{E}^{*}_{m,p}}(\mathbf{r})\mathbf{e}^*
\end{equation}

Using again the commutator between the Hamiltonian and position operator, we derive the following relation between matrix elements \cite{nisoli2016semiconductor,rosencher}: 

\begin{equation}
\mathbf{r}_{vc}= \mathbf{p}_{vc}\cdot \frac{-j\hbar}{m_0(E_v-E_c)}\simeq \mathbf{p}_{vc}\cdot \frac{j}{m_0\omega_{m^{\prime},p^{\prime}}}
\label{mat_elmt}
\end{equation}

with $m_0$ the free electron mass. $\hat{X}_{m^{\prime},p^{\prime}}$ becomes:

\begin{equation}
\hat{X}_{m^{\prime},p^{\prime}}=\sum_{m=-\infty}^{+\infty}\sum_{p=1}^{\infty}\frac{1 }{\omega_{m^{\prime},p^{\prime}} m_{0}}C_{m,p}\Phi(0)I(z)I(r,\theta)\hat{a}_{m,p}^{\dagger}
\end{equation}

with 

\begin{align}
I(z)& =\int{dz\chi_{c}(z)\chi_{v}(z)\tilde{E}_x^*(z)}\simeq \tilde{E}_x^*(z_{QW})\int{dz\chi_{c}(z)\chi_{v}(z)}	\\
I(r,\theta)& =\int{rdrd\theta F_{m^{\prime},p^{\prime}}(r,\theta)\tilde{E}^{*}_{m,p}(r,\theta)\left[\mathbf{p}_{vc} \cdot \mathbf{e}^*(r,\theta) \right]}
\label{eq:ir}
\end{align}

 $I(z)$ is practically identical to the overlap integral of the electron and hole wave function along the $z$ axis. This integral also involves the z distribution of the electromagnetic mode: this implies that the exciton can only couple to a cavity mode sharing the same parity in $z$. In our case, these z-dependent functions are symmetric: cosine function inside the disk, with exponentially decreasing tails outside. At the scale of the quantum well, one can neglect the z-variations of the electric field, thus $\tilde{E}_x^*(z)\simeq \tilde{E}_x^*(z_{QW})$. Here the QW sits in the middle of disk, hence $\tilde{E}_x^*(z_{QW})=1$ \\

After injecting the expressions in equation \eqref{ham_int2}, the interaction Hamiltonian takes the form: 

\begin{equation}
H_{I}=\sum_{m,m'=-\infty}^{+\infty}\sum_{p^{\prime},p=1}^{+\infty}\hbar g^{m,m^{\prime},p,p^{\prime}}_{cx}(\hat{a}_{m,p}\hat{d}_{m^{\prime},p^{\prime}}^{\dagger}+\hat{a}_{m,p}^{\dagger}\hat{d}_{m^{\prime},p^{\prime}})
\end{equation} 

with 

\begin{equation}
\hbar g^{m,m',p,p^{\prime}}_{cx}=\frac{e}{m_{0}}C_{m,p}\Phi(0)I(z)I(r,\theta)
\label{C1}
\end{equation}

The scalar product $\left[\mathbf{p}_{vc} \cdot \mathbf{e}^*(r,\theta) \right]$ is evaluated using standard momentum matrix elements \footnote{See R. De Oliveira, PhD Thesis, Université Paris Cité 2022, for derivation.}. The coupling parameter $g_{cx}$ can now be expressed for the disk WGM and excitonic wave function:

\begin{align}
\begin{alignedat}{2}
\hbar g^{m,m',p,p^{\prime}}_{cx} & =\frac{e}{\sqrt{2}m_{0}}C_{m,p}\Phi(0)I(z)P_x \int rdr && d\theta  F_{m^{\prime},p^{\prime}}(r,\theta)[\tilde{E}^{*}_{m,p,+}e^{j \theta} +\tilde{E}^{*}_{m,p,-}e^{- j \theta} ](r,\theta) \\
& = \frac{e}{m_{0}}\frac{C_{m,p}\Phi(0)I(z){P_x}\sqrt{2}}{\sqrt{\max\left(\varepsilon(\mathbf{r}){\left|\mathbf{E}_{m,p}(\mathbf{r})\right|}^{2}\right) }} && \int_{0}^{R}{dr\frac{\sqrt{\varepsilon(r)}}{\beta_{m,p}}E_{0,m,p}K_{m^{\prime},p^{\prime}}mJ_{m'}(\frac{x_{m^{\prime},p^{\prime}}r}{R})J_{m}(\beta_{m,p} r)}\\
& && \times  \int_{0}^{2\pi}{e^{j(m-m')\theta}d\theta} 
\end{alignedat}
\label{eq:gcx1}
\end{align}

where $P_x=\vert \mathbf{p}_{cv} \vert=\sqrt{\frac{E_P m_0}{2}}$ is a parameter linked to the tabulated Kane energy $E_P$. Transition to the second line is done using the equations \eqref{eq:field_3D_TE} and \eqref{wave_ex}. The second integral imposes $m=m'$ to obtain non-zero coupling: as required by symmetry, the azimuthal number $m$ is conserved in the coupling.  Using the definition of the oscillator strength per unit of area: 

\begin{equation}
\frac{f}{S}=\frac{2}{m_{0}\hbar \omega_{m^{\prime},p^{\prime}}}{\vert \mathbf{p}_{cv}\vert}^{2}{\vert\Phi(0)\vert}^{2}{\left|\int{dz \chi_{c}(z) \chi_{v}(z)}\right|}^{2}
\end{equation}

and assuming that $ \omega_{m^{\prime},p^{\prime}}\simeq \omega_{m,p}$ the final expression for the coupling constant is: 

\begin{equation}
\hbar g^{m,p,p^{\prime}}_{cx}= {\hbar}\sqrt{\frac{e^{2}}{2\varepsilon_{0}\varepsilon_{r}m_{0}V_{\text{eff}}^{m,p}}\frac{f}{S}}\int_{0}^{R}{dr\space C_{m,p,p^{\prime}}J_{m}\left(\frac{x_{m,p^{\prime}}r}{R}\right)J_{m}(k_{m,p}n_er)}
\label{eq:final}
\end{equation}

where $$C_{m,p,p{\prime}}=\frac{2\pi m}{[\max(J_{m-1}^2(k_{m,p}n_e r)+J_{m+1}^2(k_{m,p}n_e r))]^{1/2}}\frac{1}{R\sqrt{\pi}k_{m,p}n_e\vert J_{m-1}(x_{m,p^\prime}) \vert} $$ is a dimensionless parameter originating from the normalization of both excitonic and photonic modes. To evaluate this last parameter we used the fact that the electric field is taking its maximum value inside the disk resonator. The final result takes the form of an exciton-photon overlap integral, which has the dimension of a length, multiplied by a prefactor taking into account the properties of the quantum well (oscillator strength) and the WGM (mode volume), which has the dimension of an energy divided by a length. Interestingly, there is no strict selection rule on the radial quantum numbers $p$ and $p^{\prime}$: the overlap can be non-zero even if $p\neq p^{\prime}$. This said, for $m=m'$ and $p=p^{\prime}$ the exciton wave function and the electromagnetic mode distribution have similar geometry, governed by the Bessel function, and the overlap integral is maximized. \\

\begin{figure}[h!]
\centering
\includegraphics[scale=0.9]{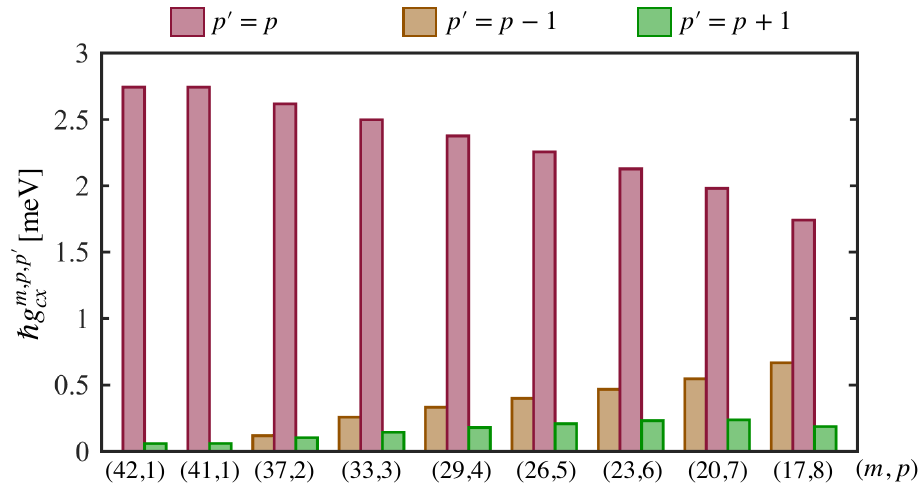}
\caption{Calculation of the coupling energy $\hbar g^{m,p,p^{\prime}}_{cx}$  for a single QW located in the middle of the disk considered in our experiments, for different couples of azimuthal and radial numbers ($m,p$). Three situations are considered: $p^{\prime} = p$ (red), $p^{\prime}=p-1$ (brown) and $p^{\prime}=p+1$ (green).}
\label{fig:GCX}
\end{figure}

Figure \ref{fig:GCX} shows the results obtained for the computation of $\hbar g^{m,p,p^{\prime}}_{cx}$ for different couples ($m,p$): all these modes lie in the spectral vicinity of the exciton energy at cryogenic temperature for our experimental device. The coupling is very sensitive to the radial number $p$ and seems to be less impacted by a change of the azimuthal number $m$. Despite the absence of a strict selection rule for the radial number, the coupling is still greatly reduced when $p^{\prime} \neq p$, as expected from the shape of the overlap integral. However, for increasing $p$, this behavior becomes less pronounced, and non-negligible coupling can exist between a photonic and excitonic mode of distinct radial number ($p^{\prime} = p \pm 1$). For low $p$ values, the two Bessel functions are almost in perfect concordance ($\beta_{m,p}\simeq x_{m,p}/R$), and the overlap integral behave almost as a Dirac function.\\

The formula \eqref{eq:final} derived in the case of the a single QW generalizes to the MQW case \footnote{See R. De Oliveira, PhD Thesis, Université Paris Cité 2022, for derivation.} :

\begin{equation}
\hbar g_{cx}^{m,p,p^{\prime}}(N_{QW})=\sqrt{\sum^{{N}_{QW}}_{i=1}{\frac{\eta_i^2}{N_{QW}}}} \cdot \sqrt{N_{QW}}\hbar g_{cx}^{m,p,p^{\prime}}=\sqrt{\sum^{{N}_{QW}}_{i=1}{\frac{1}{N_{QW}}\frac{E_x^2(z_{QW,i})}{E_x^2(0)}}} \cdot\sqrt{N_{QW}}\hbar g_{cx}^{m,p,p^{\prime}}
\end{equation}

with $N_{QW}$ the number of QW in the structure. The coupling of the QW to the electromagnetic field depends on its position along the z-axis: its contribution is weighted by the quantity $\eta$. For five QWs, a structure that we use in our sample, the enhanced coupling is equal to $1.89\times g_{cx}^{m,p,p^{\prime}}$, instead of the idealized factor $\sqrt{5}\simeq 2.23$ that one may have in mind. Eventually, the Rabi coupling is $2\hbar g_{cx}$.


\end{document}